\documentclass[preprint]{ptephy_v1}

\preprintnumber{J-PARC-TH-0165} 

\usepackage{graphicx, fancybox}
\usepackage{amsmath, amssymb, mathrsfs, bm}
\usepackage[T1]{fontenc} 
\usepackage[table]{xcolor}
\usepackage{ulem}
\usepackage{hyperref}

\usepackage{slashbox,multirow}

\hypersetup{
    pdffitwindow=true,      
    pdfnewwindow=true,      
    bookmarksnumbered=true,
    colorlinks=true,       
    linkcolor=red,          
    citecolor=darkgreen,        
    filecolor=magenta,      
    urlcolor=cyan,           
    menucolor=blue,
}
\definecolor{darkgreen}{rgb}{0,.7,0}

\numberwithin{equation}{section}

\begin{document}

\title{A Study of Stress-Tensor Distribution around Flux Tube \\ in Abelian-Higgs Model}

\author[1]{Ryosuke Yanagihara\thanks{yanagihara@kern.phys.sci.osaka-u.ac.jp}}
\affil{Department of Physics, Osaka University,
Toyonaka, Osaka 560-0043, Japan}

\author[1, 2]{Masakiyo Kitazawa\thanks{kitazawa@phys.sci.osaka-u.ac.jp}}
\affil{J-PARC Branch, KEK Theory Center,
Institute of Particle and Nuclear Studies, KEK,
203-1, Shirakata, Tokai, Ibaraki, 319-1106, Japan}

\begin{abstract}
 We study the stress-tensor distribution around the flux tube 
 in static quark and anti-quark systems based on the momentum
 conservation and the Abelian-Higgs (AH) model.
 We first investigate constraints on the stress-tensor distribution
 from the momentum conservation
 and show that the effect of boundaries plays a crucial role
 to describe the structure of the flux tube 
 in SU(3) Yang-Mills theory which has measured recently on the lattice.
 We then study the distributions of the stress tensor and energy density
 around the magnetic vortex with and without boundaries
 in the AH model, and compare them with the distributions 
 in SU(3) Yang-Mills theory based on the dual superconductor picture.
 It is shown that a wide parameter range of the AH model is excluded
 by a comparison with the lattice results in terms of the stress tensor.
\end{abstract}

\subjectindex{}

\maketitle

\section{Introduction}
\label{sec:intro}

The energy-momentum tensor (EMT) 
$\mathcal{T}_{\mu\nu}(x)$
is an important observable in various fields in physics including 
gravitational theory, hydrodynamics, and elastic body.
Among the components of EMT, its spatial part, which is related to 
the stress tensor $\sigma_{ij}$ as $\sigma_{ij}=-\mathcal{T}_{ij}$
with $i,j=1,2,3$, 
is a fundamental observable related to force acting on a surface.
In field theory, the stress tensor represents 
distortion of fields induced by external sources~\cite{Landau}.
For example, in Maxwell theory local propagation of a 
Coulomb interaction between charges is 
characterized by the Maxwell stress, which is 
the spatial component of the EMT in this theory,
$\mathcal{T}_{\mu\nu}=
F_{\mu\rho} F^{\rho}_{\; \nu} - (1/4)\delta_{\mu\nu} F_{\rho\sigma}F^{\rho\sigma}$
with the field strength $F_{\mu\nu}$~\cite{Landau}.
The stress tensor in non-Abelian gauge theories including QCD is
even more important
because this observable characterizes the structure of the
non-Abelian fields with external sources in a gauge invariant
manner.
Recently, the analysis of the stress tensor has been performed
in various systems described by the strong interaction,
such as the static-quark systems~\cite{Yanagihara:2018qqg},
hadrons~\cite{Kawamura:2013wfa,Kumano:2017lhr,Burkert:2018bqq,
  Pagels:1966zza,Polyakov:2018zvc,Shanahan:2018pib,Shanahan:2018nnv},
and thermal system having a pressure anisotropy~\cite{Kitazawa:2019otp}.

In Ref.~\cite{Yanagihara:2018qqg},
the stress-tensor distribution in static quark 
and an anti-quark ($Q\bar{Q}$) systems in SU(3) Yang-Mills (YM) theory
has been measured in the numerical simulation of lattice gauge theory.
In this study, the analysis of the stress tensor on the lattice is realized
with the EMT operator~\cite{Suzuki:2013gza,Asakawa:2013laa,Makino:2014taa,
Kitazawa:2016dsl,Taniguchi:2016ofw}
constructed via the gradient flow~\cite{Narayanan:2006rf,Luscher:2010iy,Luscher:2011bx}.
Through the analysis of the principal directions and eigenvalues of
the stress tensor, the formation of the flux tube is revealed 
in terms of the gauge invariant observable.
Before this study, 
the spatial structure of the flux tube between $Q\bar{Q}$ had been
investigated using the color electric field and
action density~\cite{DiGiacomo:1990hc,Bali:1994de,Michael:1995pv,Green:1996be,Gliozzi:2010zv,Meyer:2010tw,Cea:2012qw,Cardoso:2013lla,Cea:2015wjd,Cea:2017ocq} 
(see also the reviews~\cite{Bali:2000gf,Greensite,Kondo:2014sta}).
Compared with these previous studies,
the use of EMT and especially the stress tensor has several advantages.
First, EMT is gauge invariant and an observable having a definite physical
meaning related to energy density and force acting on a surface.
For example, principal directions of the stress tensor 
serve as a gauge invariant definition of the direction of ``line of force''
in non-Abelian theories~\cite{Yanagihara:2018qqg}.
Moreover, EMT is a renormalization-group invariant quantity and its
absolute value has an unambiguous meaning.
Second, as EMT is a second-rank tensor having many channels compared
with a vector field, it provides us with more detailed 
information on the system than the color electric field.
In fact, in Ref.~\cite{Yanagihara:2018qqg}
it was found that the eigenvalues of EMT on the mid-plane
between $Q\bar{Q}$ shows nontrivial degeneracies and separations.

In the present study, motivated by the numerical results in
Ref.~\cite{Yanagihara:2018qqg}
we explore the distribution of EMT in 
the $Q\bar{Q}$ system using conservation laws and a specific model%
\footnote{
  Preliminary results of the present paper are reported
  in Refs.~\cite{Yanagihara:JPS2019} and \cite{Kitazawa:2019uel}.
  See, also Ref.~\cite{Nishino:2019bzb}.
}.
We first discuss constraints on the EMT distribution from the
momentum conservation.
We show that the transverse structure of the eigenvalues of EMT
must have a separation for an infinitely-long flux tube
having a translational invariance.
This property is qualitatively inconsistent with the lattice results
in Ref.~\cite{Yanagihara:2018qqg}.
The momentum conservation thus leads to a conclusion that the effect of
boundaries of the flux tube is crucial to describe the lattice results.

We then employ the Abelian-Higgs (AH) model and study
the EMT distribution around the magnetic vortex with and without
boundaries.
The AH model is a relativistically generalized version of 
the Ginzburg-Landau (GL) model for superconductivity.
According to the dual superconductor picture~\cite{Nambu:1974zg,tHooft:1975krp,Mandelstam:1974pi,Nielsen:1973cs}, 
the dual of the AH model is regarded as a phenomenological
model of low energy QCD; attempts to derive the dual AH model from QCD
have been discussed 
in the literature~\cite{Suzuki:1988yq,Ball:1987cf,Maedan:1989ju,Kodama:1997zc}
based on the Abelian dominance~\cite{tHooft:1981bkw} and
the monopole 
condensation~\cite{Kronfeld:1987vd,Kronfeld:1987ri,Maedan:1988yi}.
In this picture,
magnetic monopoles and the magnetic vortex between the monopoles
in the AH model correspond to the color charges and 
the flux tube in YM theory~\cite{Nielsen:1973cs}, respectively.
Based on this picture, 
the vortex solution in the AH model has been compared with 
numerical results on the flux tube in 
YM theory~\cite{Suganuma:1993ps,Sasaki:1994sa,Ichie:1994eg,Monden:1997hb,Koma:1999sm,Koma:2003gq,Oxman:2017boz,Oxman:2018dzp}.
In these studies, however, action density and/or 
field strength of the color gauge field in a specific gauge
have been used for observables to make a comparison.

In the present study, we calculate the spatial distribution of EMT
around the magnetic vortex in the AH model,
and compare it with the lattice result in Ref.~\cite{Yanagihara:2018qqg}.
We demonstrate that the stress-tensor distribution
around the infinitely-long magnetic vortex is qualitatively inconsistent
with the lattice result, as anticipated from the momentum conservation.
We then analyze the magnetic vortex with a finite length, and 
show that a wide parameter range of the AH model with a standard potential
cannot reproduce the lattice result 
in Ref.~\cite{Yanagihara:2018qqg} simultaneously.

This paper is organized as follows.
In Sec.~\ref{sec:stress}, we discuss general properties of the
stress-tensor distribution around the flux tube which are obtained
only from the momentum conservation and cylindrical symmetry.
We then employ the AH model in Sec.~\ref{sec:ah}, and
discuss the magnetic vortex in this model
with and without boundaries in Sec.~\ref{sec:vortex}.
In Sec.~\ref{sec:result},
we discuss the numerical results on the stress-tensor distribution
around the magnetic vortex.
The final section is devoted to a short summary.
Some analytic properties of the vortex solution in the AH model
is summarized in Appendix~\ref{sec:analytic}.

Throughout this paper we consider $3+1$ dimensional Minkowski space
with the metric $g^{\mu\nu}={\rm diag}(1,-1,-1,-1)$ 
with $\mu,\nu=0,1,2,3$.

\section{Stress tensor and momentum conservation}
\label{sec:stress}

In this section, 
we summarize general properties of the EMT distribution around the flux tube
which do not depend on a specific model.
In particular, we discuss constraints 
from the momentum conservation, and show that 
the lattice results in Ref.~\cite{Yanagihara:2018qqg} are qualitatively
inconsistent with the flux tube with an infinite length.

\subsection{Stress tensor}
\label{sec:stress-sub}

The stress tensor is related to the spatial components of EMT 
as~\cite{Landau}
\begin{align}
 \sigma_{ij}=-\mathcal{T}_{ij} \quad (i,j=1,2,3).
\end{align}
Force per unit area ${\cal F}_i$
acting on a surface with the normal vector $n_i$ is represented
in terms of the stress tensor as
\begin{eqnarray}
 {\cal F}_i =  \sigma_{ij}n_j =  - {\cal T}_{ij} n_j.
 \label{eq:F=Tn}
\end{eqnarray} 
From Eq.~(\ref{eq:F=Tn}) one sees that the force
and the normal vector are parallel only for the local principal
axes obtained by solving the eigenvalue equations
\begin{align}
 {\cal T}_{ij}n_j^{(k)}=\lambda_k n_i^{(k)} \quad (k=1,2,3).
 \label{eq:EV}
\end{align}
The strength of the force per unit area
along $n_i^{(k)}$ is given by the eigenvalue $\lambda_k $.
Neighboring volume elements separated by a surface with the normal
vector $n_i^{(k)}$ are pulling (pushing) with each other
for $\lambda_k <0 $ ($\lambda_k >0 $) on the surface.
As $\sigma_{ij}$ is a symmetric tensor, three principal axes $n_i^{(k)}$
are orthogonal with each other.

Let us see two examples of EMT and stress tensor.
First, in a thermal medium with an infinite volume EMT is given by 
\begin{align}
 {\cal T}_{\mu\nu} = {\rm diag}( \varepsilon, P , P , P ),
\end{align}
with energy density $\varepsilon$ and pressure $P>0$.
As the stress tensor reads $\sigma_{ij} = -P\delta_{ij}$, 
force acting on a surface element is always
perpendicular to the surface.
The sign of the eigenvalues means that volume elements are pushing
with each other with pressure $P$.
Second, in Maxwell theory for electromagnetism EMT is given by 
\begin{align}
 \mathcal{T}_{\mu\nu}
 =  F_{\mu\rho} F^{\rho}_{\nu} - \frac{1}{4} \delta_{\mu\nu}
 F_{\rho\sigma} F^{\rho\sigma},
 \label{eq:emt_maxwell}
\end{align}
with the field strength $F_{\mu\nu}$.
When a static electric or magnetic field along the $z$ direction is applied
as $\vec{E}=(0,0,E)$ or $\vec{B}=(0,0,B)$, 
one has
\begin{align}
 {\cal T}_{\mu\nu}
 = \frac12 {\rm diag}( E^2 , E^2 , E^2 , -E^2 ),
 \quad\mbox{or}\quad
 {\cal T}_{\mu\nu}
 = \frac12 {\rm diag}( B^2 , B^2 , B^2 , -B^2 ), 
 \label{eq:T=diagE}
\end{align}
respectively.
Equation~(\ref{eq:T=diagE}) shows that volume elements are pulling
with each other along the direction of $\vec{E}$, while volume elements 
are pushing with each other along directions perpendicular 
to $\vec{E}$~\cite{Landau}.
In Eq.~(\ref{eq:T=diagE}), all absolute values of the eigenvalues of
${\cal T}_{\mu\nu}$ are identical, and 
the principal axis associated with the negative eigenvalue $\lambda_i$
is parallel to the field $\vec{E}$ or $\vec{B}$.
This principal axis thus corresponds to the direction
of the line of force in Maxwell theory.

In a static system, the momentum conservation implies
\begin{align}
 \partial_i \mathcal{T}^{ij} = 0.
 \label{eq:conservation}
\end{align}
By taking a volume integral of Eq.~(\ref{eq:conservation}) on a volume $V$
without external charges
and using the Gauss theorem, one obtains 
\begin{align}
 \int_V dV\, \partial_i \mathcal{T}^{ij} 
 = \int_S dS_i\, \mathcal{T}^{ij} = 0,
 \label{eq:dS}
\end{align}
where $S$ is the surface of $V$ with the outgoing surface vector.
Since $dS_i \mathcal{T}^{ij}$ is the $j$-th component of force acting
on the surface element $dS_i$, Eq.~(\ref{eq:dS}) represents the
equilibrium of force acting on $V$ through its surface.
When there exists a test charge in volume $V$, 
force $\vec{F}$ acting on the test charge is related to the
surface integral as ${F}_i =- \int_S {\cal T}_{ij} dS_j$.

\subsection{Cylindrical coordinate system}
\label{sec:cylindrical}

In the analysis of the flux tube or magnetic vortex,
it is convenient to employ the cylindrical coordinate system
$(r,\theta,z)$ with $r=\sqrt{x^2+y^2}$ and $\theta=\tan^{-1} (y/x)$
with $0\le\theta<2\pi$
because of the rotational symmetry around an axis.
The components of EMT in this coordinate system are given by
\begin{align}
 \mathcal{T}_{\gamma\gamma'}(r,z)=
 (e_\gamma)_\mu \mathcal{T}^{\mu\nu} (e_{\gamma'})_\nu , 
 \label{eq:Tgamma}
\end{align}
with $\gamma,\gamma' = 0, r, \theta, z$ and $e_\gamma$
denotes a unit vector along the direction of the $\gamma$ axis
in the Minkowski space.

The momentum conservation Eq.~(\ref{eq:conservation}) in terms of 
$\mathcal{T}_{\gamma\gamma'}(r,z)$ is given by
\begin{align}
 &\frac{1}{r}\partial_r(r\mathcal{T}_{rr})
 -\frac{\mathcal{T}_{\theta\theta}}{r}
 +\partial_z \mathcal{T}_{rz}=0,
 \label{eq:con-law-fin}
 \\
 &\partial_\theta \mathcal{T}_{\theta\theta}=0,
 \\
 &\frac{1}{r}\partial_r(r\mathcal{T}_{rz})+\partial_z \mathcal{T}_{zz}=0,
  \label{eq:con-law-fin2}
\end{align}
which represents the equilibrium of force acting on an infinitesimal
volume element along the $r,\theta,z$ directions, respectively.

When a system is translationally symmetric along $z$ direction
in addition to the rotational symmetry,
the components of EMT is given by functions of $r$ as
${\cal T}_{\gamma\gamma'}={\cal T}_{\gamma\gamma'}(r)$, and 
$z$ derivatives in Eqs.~(\ref{eq:con-law-fin})--(\ref{eq:con-law-fin2}) 
vanish.
Only a non-trivial constraint from the momentum conservation
is then obtained from Eq.~(\ref{eq:con-law-fin}) as
\begin{align}
 \partial_r(r\mathcal{T}_{rr}(r)) = \mathcal{T}_{\theta\theta}(r).
 \label{eq:cons-inf}
\end{align}
By rewriting Eq.~(\ref{eq:cons-inf}) as
$r \partial_r(\mathcal{T}_{rr}(r))
= \mathcal{T}_{\theta\theta}(r) - \mathcal{T}_{rr}(r)$ it is easy to show
that $\mathcal{T}_{rr}(r)$
and $\mathcal{T}_{\theta\theta}(r)$ behave differently as functions of $r$
except for the case 
$\mathcal{T}_{rr}(r) = \mathcal{T}_{\theta\theta}(r) = 0$
provided that $\mathcal{T}_{rr}(r)\to0$ 
in the $r\to\infty$ limit.
Moreover, by integrating out both sides of Eq.~(\ref{eq:cons-inf}) and 
assuming $r\mathcal{T}_{rr}(r)|_{r\to\infty} = r\mathcal{T}_{rr}(r)|_{r\to0} = 0$ 
one obtains 
\begin{align}
 \int_0^\infty dr\, {\cal T}_{\theta\theta}(r) = 0,
 \label{eq:theta=0}
\end{align}
which shows that ${\cal T}_{\theta\theta}(r)$ as a function of $r$
must change sign at least once so that its integral vanishes.

\subsection{Lattice results on flux tube in SU(3) YM theory}
\label{sec:lattice}

\begin{figure}
 \includegraphics[width=0.32\textwidth]{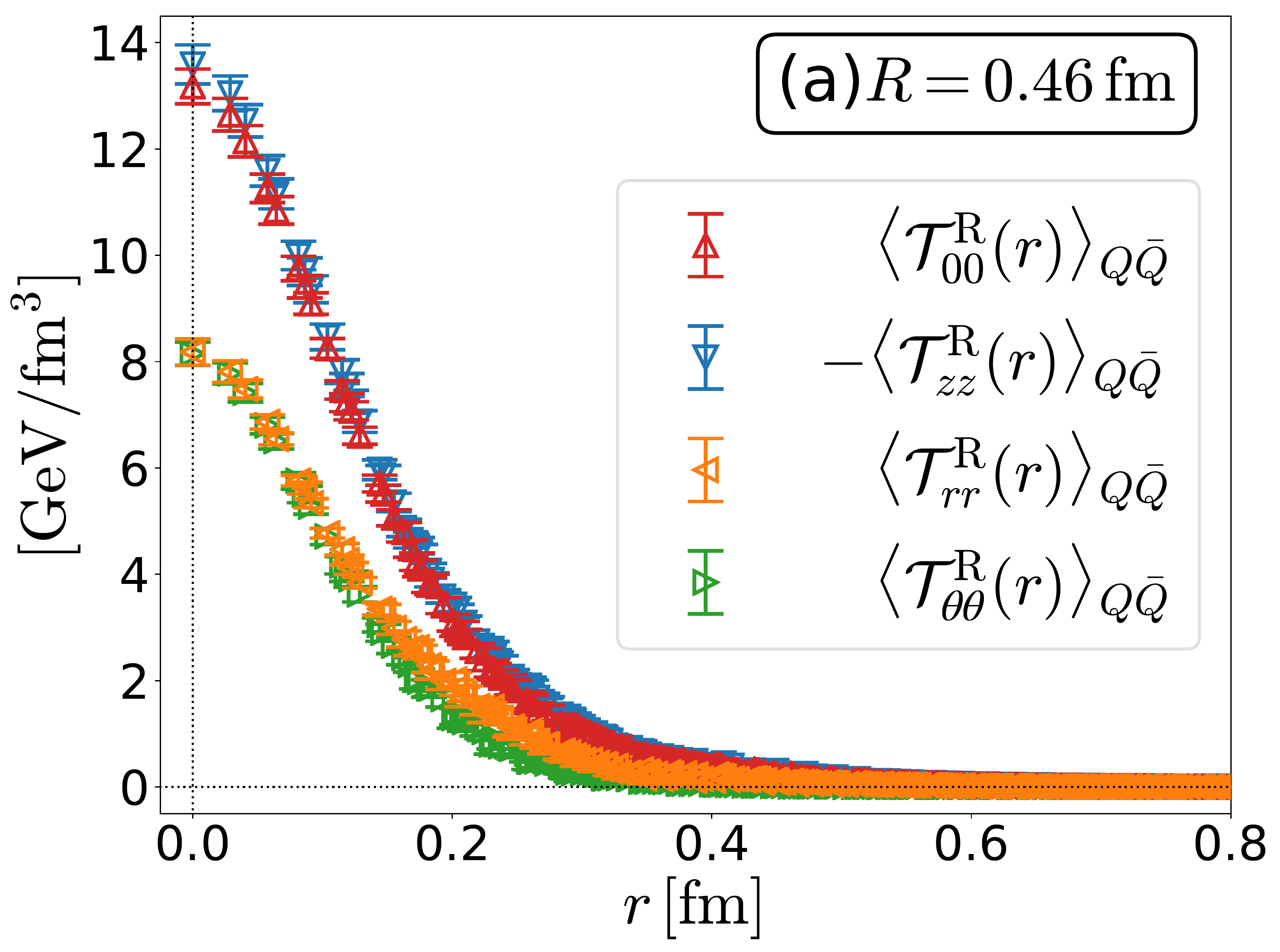}
 \includegraphics[width=0.32\textwidth]{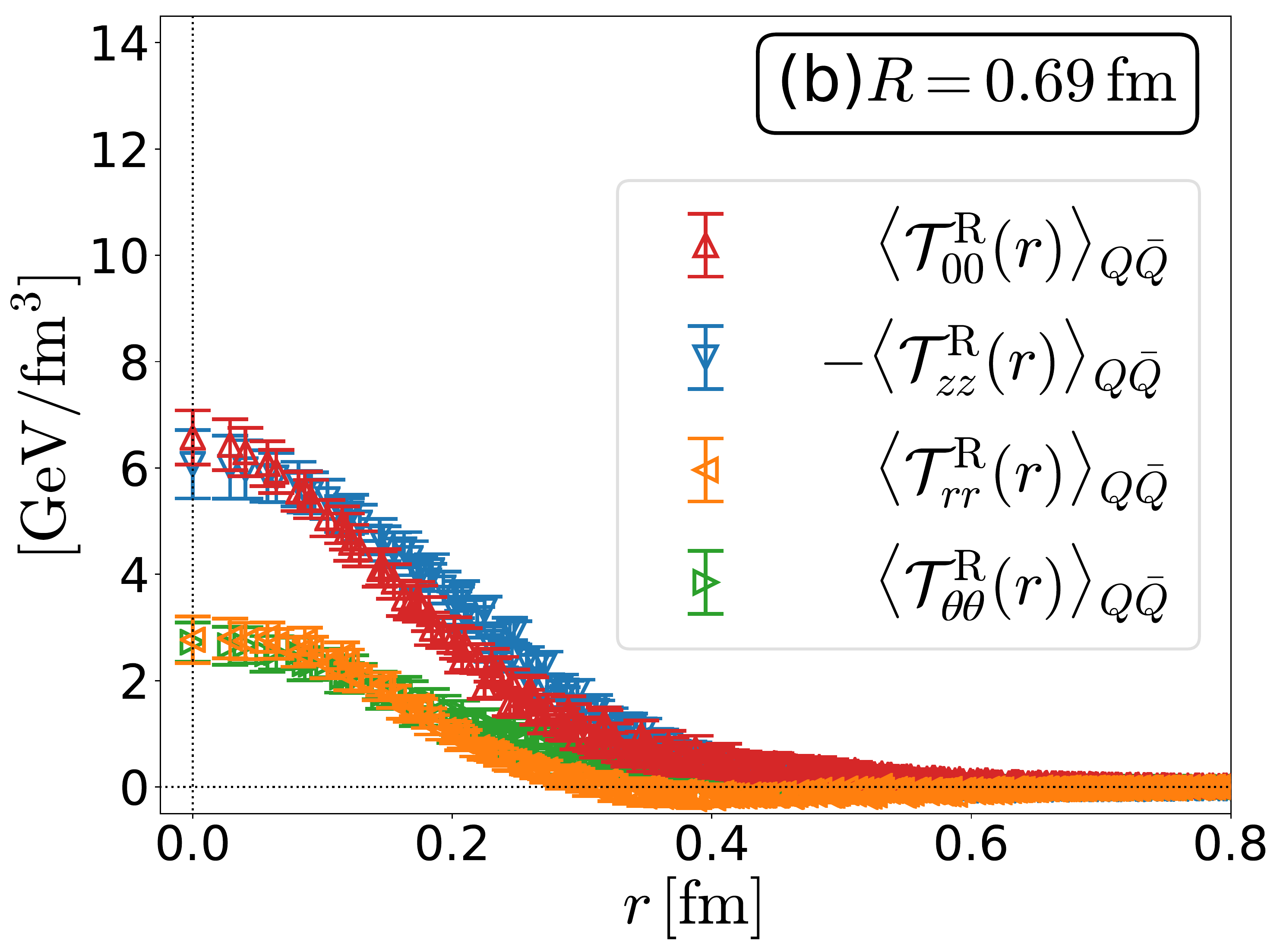}
 \includegraphics[width=0.32\textwidth]{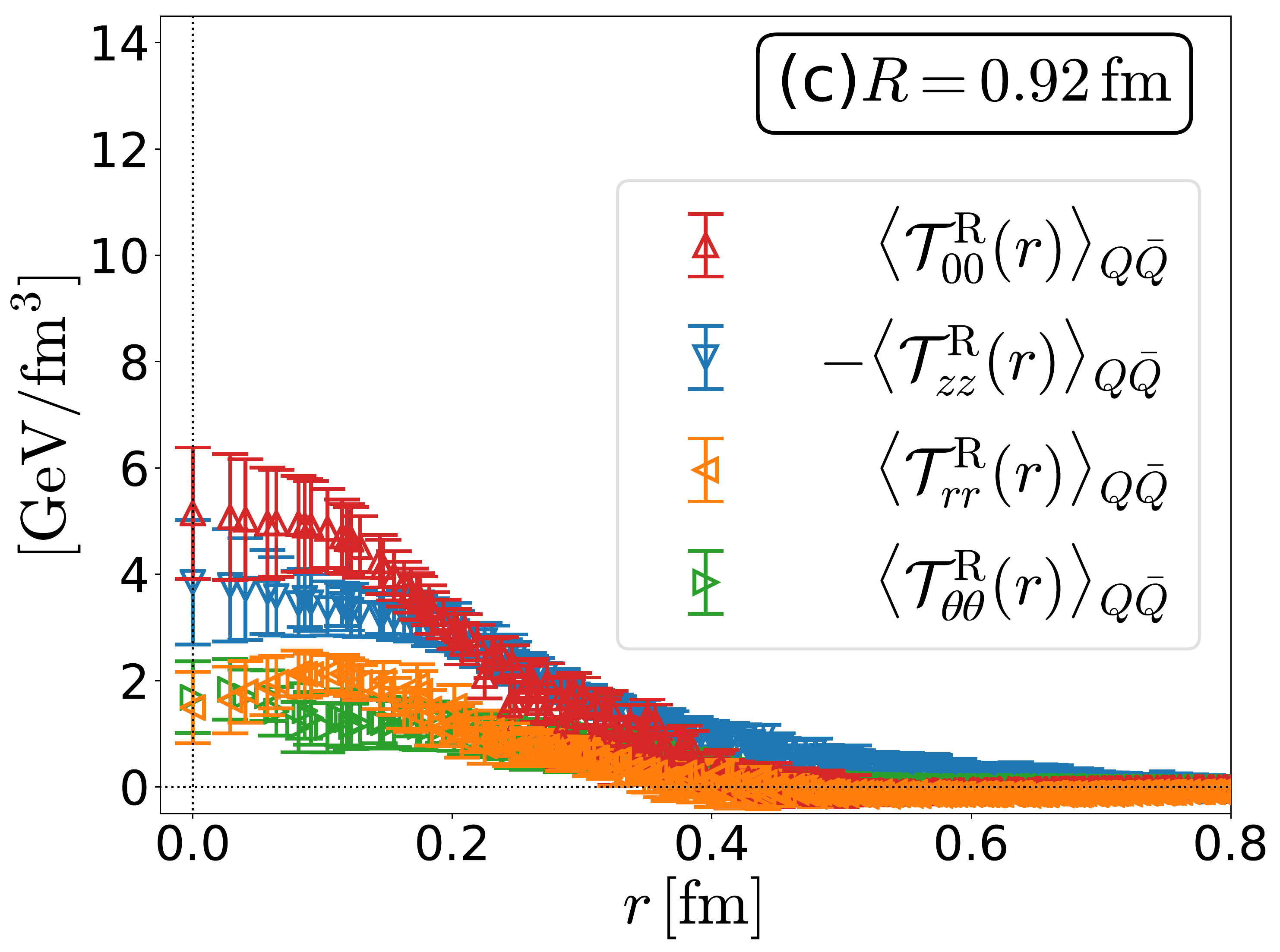}
 \caption{
 Distribution of EMT on the mid-plane of the flux tube
 obtained from the lattice numerical
 simulation in SU(3) YM theory~\cite{Yanagihara:2018qqg}.
 The lengths of the flux tube
 are $R=0.46,$~$0.69,$ and $0.92$~fm for (a), (b), and (c), respectively.
 }
 \label{fig:mid_3lat}
\end{figure}

Now let us inspect the components of EMT on the mid-plane of
the flux tube in SU(3) YM theory
obtained on the lattice in Ref.~\cite{Yanagihara:2018qqg}.
In Fig.~\ref{fig:mid_3lat} we show the expectation values of 
${\cal T}_{00}(r)$, 
${\cal T}_{zz}(r)$, ${\cal T}_{rr}(r)$, 
and ${\cal T}_{\theta\theta}(r)$ on the mid-plane in the $Q\bar{Q}$ system
as functions of $r$ for three $Q\bar{Q}$ distances,
$R=0.46$, $0.69$, and $0.92$~fm.
The figure shows that ${\cal T}_{rr}(r)$ and ${\cal T}_{\theta\theta}(r)$
are degenerated within statistics for all $Q\bar{Q}$ distances.
Moreover, the result suggests 
$\int_0^\infty dr {\cal T}_{\theta\theta}(r)>0$.
These properties do not agree with Eqs.~(\ref{eq:cons-inf}) and
(\ref{eq:theta=0}) obtained by assuming the translational invariance.
Therefore, the result in Fig.~\ref{fig:mid_3lat} shows
that the assumption of translational invariance is not applicable
to the flux tube in SU(3) YM theory even on the mid-plane with $R=0.92$~fm.

\section{Abelian-Higgs model}
\label{sec:ah}

\subsection{Model}
\label{sec:model}

Now we employ the Abelian-Higgs (AH) model and
investigate the EMT distribution around the magnetic vortex
with and without boundaries.
Our starting point is the AH Lagrangian
in four-dimensional Minkowski space:
\begin{align}
 \mathcal{L}_{\mathrm{AH}}
 &=-\frac{1}{4g^2}(\partial_\mu A_\nu(x)-\partial_\nu A_\mu(x))^2
 +|(\partial_\mu +iA_\mu(x))\chi(x)|^2-\lambda(|\chi(x)|^2-v^2)^2
 \nonumber
 \\
 &=-\frac{1}{4g^2}F_{\mu\nu}^2(x)
 +|D_\mu\chi(x)|^2 - V(|\chi(x)|),
 \label{eq:L}
\end{align}
where $F_{\mu\nu}(x)=\partial_\mu A_\nu(x)-\partial_\nu A_\mu(x)$
and $D_\mu=\partial_\mu+iA_\mu(x)$ are
the field strength tensor and
the covariant derivative, respectively,
with the Abelian gauge field $A_\mu(x)$
and the complex scalar field $\chi(x)$.
The last term $V(|\chi(x)|)=\lambda(|\chi(x)|^2-v^2)^2$
represents the Higgs potential which induces the
scalar condensation.
The AH model has three parameters, $g$, $\lambda$, and $v$.
$g$ is the gauge coupling constant, and 
$v$ is the vacuum expectation value of $\chi(x)$.
The EMT in the AH model is obtained as a Noether current
of the translational symmetry as~\cite{Bogomolnyi:1976,deVega:1976xbp}
\begin{align}
 \mathcal{T}_{\mu\nu}(x)
 =(D_\mu\chi(x))^*(D_\nu\chi(x))+(D_\nu\chi(x))^*(D_\mu\chi(x))
 -\frac{1}{g^2}g^{\rho\sigma}F_{\mu\rho}(x)F_{\nu\sigma}(x)
 -g_{\mu\nu}\mathcal{L}.
 \label{eq:emt-master}
\end{align}

The AH model has two characteristic length scales;
the correlation lengths of the scalar and gauge fields~\cite{Fetter}, 
\begin{align}
 \xi_\chi=\frac1{2\sqrt{\lambda}v},\quad \xi_A=\frac1{\sqrt{2}gv},
 \label{eq:corr}
\end{align}
respectively.
The ratio of these two scales characterizes 
the Ginzburg-Landau (GL) parameter $\kappa$, 
\begin{align}
 \kappa=\frac{\xi_\chi}{\sqrt{2} \xi_A}=\frac{\sqrt{\lambda}}{g}.
 \label{eq:kappa}
\end{align}
In the context of superconductivity, the GL parameter classifies
the type of superconductor into 
type-I ($\kappa<1/\sqrt{2}$) or type-II ($\kappa>1/\sqrt{2}$).
In the type-I superconductor, 
the interaction between vortices is attractive so that 
a system of two and more individual vortices is unstable
for the formation of a single vortex with a multiple winding number.
In contrast, 
the interaction between vortices is repulsive 
in the type-II superconductor~\cite{Fetter}.
The boundary value $\kappa = 1/\sqrt{2}$ is called 
the Bogomol'nyi bound~\cite{Bogomolnyi:1976}.

The AH Lagrangian 
is invariant under the U(1) gauge transformation
defined by 
\begin{align}
 \chi(x)\rightarrow\chi(x)e^{i\alpha(x)},\quad
 \chi^*(x)\rightarrow\chi^*(x)e^{-i\alpha(x)},\quad
 A_\mu(x)\rightarrow A_\mu(x)-\partial_\mu\alpha(x),
\end{align}
with a gauge function $\alpha(x)$.
The complex scalar field $\chi(x)$ is written as
\begin{align}
 \chi(x)=\phi(x)e^{if(x)}, 
\end{align}
with a real scalar field $\phi(x)$ and a phase function $f(x)$.
In the unitary gauge defined by $\alpha(x)=-f(x)$, 
we have the gauge-fixed AH Lagrangian,
\begin{align}
 \mathcal{L}_{\mathrm{AH}}
 =-\frac{1}{4g^2}(\partial_\mu A'_\nu(x)-\partial_\nu A'_\mu(x))^2
 +|(\partial_\mu+iA'_\mu(x))\phi(x)|^2-\lambda(\phi(x)^2-v^2)^2
 \label{eq:lagrangian},
\end{align}
with $ A'_\mu(x)=A_\mu(x)+\partial_\mu f(x)$.

\subsection{Energy-momentum tensor in cylindrical coordinate system}
\label{sec:EMT:AH}

In this study we investigate the spatial distribution of EMT
around the magnetic vortex with and without boundaries.
As these systems possess rotational symmetry around the vortex,
we employ the cylindrical coordinate system.
For the vortex with length $R$,
we suppose that the boundaries are at $(r,z)=(0,\pm R/2)$.
From the rotational symmetry the scalar field is given by a
functions of $r$ and $z$ as 
\begin{align}
 \phi=\phi(r,z),
\end{align}
and it is possible to represent the gauge field as
\begin{align}
 \vec{A}(r,z)=\frac{\tilde{A}(r,z)}{r}\vec{e}_\theta.
\end{align}
with a scalar function $\tilde{A}(r,z)$
and the three-dimensional unit vector $\vec{e}_\theta$ 
in the direction of the $\theta$ axis.

The non-vanishing components of EMT in the cylindrical coordinate system,
Eq.~(\ref{eq:Tgamma}), are calculated to be
\begin{align}
 &\mathcal{T}_{00}(r,z)=
 \frac{1}{2g^2r^2}
 \Bigl((\partial_r\tilde{A})^2 + (\partial_z\tilde{A})^2\Bigr)
 +\Bigl((\partial_r\phi)^2 + (\partial_z\phi)^2\Bigr)
 +\frac{\phi^2\tilde{A}^2}{r^2}
 +\lambda(\phi^2-v^2)^2, \label{eq:emt-0}
 \\
 &\mathcal{T}_{rr}(r,z)=
 \frac{1}{2g^2r^2}
 \Bigl((\partial_r\tilde{A})^2 - (\partial_z\tilde{A})^2\Bigr)
 +\Bigl((\partial_r\phi)^2 - (\partial_z\phi)^2\Bigr)
 -\frac{\phi^2\tilde{A}^2}{r^2}
 -\lambda(\phi^2-v^2)^2, \label{eq:emt-r}
 \\
 &\mathcal{T}_{\theta\theta}(r,z)=
 \frac{1}{2g^2r^2}
 \Bigl((\partial_r\tilde{A})^2 + (\partial_z\tilde{A})^2\Bigr)
 -\Bigl((\partial_r\phi)^2 + (\partial_z\phi)^2\Bigr)
 +\frac{\phi^2\tilde{A}^2}{r^2}
 -\lambda(\phi^2-v^2)^2, \label{eq:emt-theta}
 \\
 &\mathcal{T}_{zz}(r,z)=
 \frac{-1}{2g^2r^2}
 \Bigl((\partial_r\tilde{A})^2 - (\partial_z\tilde{A})^2\Bigr)
 -\Bigl((\partial_r\phi)^2 - (\partial_z\phi)^2\Bigr)
 -\frac{\phi^2\tilde{A}^2}{r^2}
 -\lambda(\phi^2-v^2)^2, \label{eq:emt-z}
 \\
 &\mathcal{T}_{rz}(r,z)=
 \frac{1}{g^2r^2}(\partial_r \tilde{A})(\partial_z \tilde{A})
 +2(\partial_r\phi)(\partial_z\phi).
 \label{eq:emt-rz}
\end{align}

On the mid-plane at $z=0$, 
$\mathcal{T}_{rz}(r,z=0)$ vanishes from the symmetric reason
and EMT is diagonalized as 
\begin{align}
 \mathcal{T}_{\gamma\gamma'}(r)
 =\mathrm{diag}(\mathcal{T}_{00}(r), \mathcal{T}_{rr}(r),
 \mathcal{T}_{\theta\theta}(r), \mathcal{T}_{zz}(r) ),
 \label{eq:Tgamma=diag}
\end{align}
where the argument $z=0$ in $\mathcal{T}_{\gamma\gamma'}(r,z)$
is abbreviated for notational simplicity.
Each component is given by
\begin{align}
 \mathcal{T}_{00}(r)=-\mathcal{T}_{zz}(r)=&
 \frac{1}{2g^2r^2}
 (\partial_r\tilde{A})^2
 +(\partial_r\phi)^2
 +\frac{\phi^2\tilde{A}^2}{r^2}
 +\lambda(\phi^2-v^2)^2 ,
 \label{eq:emt-0z-fin}
 \\
 \mathcal{T}_{rr}(r)=&
 \frac{1}{2g^2r^2}
 (\partial_r\tilde{A})^2
 +(\partial_r\phi)^2
 -\frac{\phi^2\tilde{A}^2}{r^2}
 -\lambda(\phi^2-v^2)^2 ,
 \label{eq:emt-r-fin}
 \\
 \mathcal{T}_{\theta\theta}(r)=&
 \frac{1}{2g^2r^2}
 (\partial_r\tilde{A})^2
 -(\partial_r\phi)^2
 +\frac{\phi^2\tilde{A}^2}{r^2}
 -\lambda(\phi^2-v^2)^2 ,
 \label{eq:emt-theta-fin}
\end{align}
where we used the fact that terms including $\partial_z$ vanish
on the mid-plane.

Equations~(\ref{eq:emt-0z-fin})--(\ref{eq:emt-theta-fin})
tell us several notable features.
First,
the absolute values of $\mathcal{T}_{00}(r)$ and
$\mathcal{T}_{zz}(r)$ on the mid-plane always degenerate in the AH model.
Second, 
the first term of Eqs.~(\ref{eq:emt-0z-fin})--(\ref{eq:emt-theta-fin})
corresponds to the Maxwell stress Eq.~(\ref{eq:emt_maxwell}).
On the mid-plane, direction of the magnetic field is along the $z$ axis.
As a result, as in Eq.~(\ref{eq:T=diagE}) this term gives a 
negative (positive) contribution to 
$\mathcal{T}_{zz}(r)$ ($\mathcal{T}_{rr}(r)$ 
and $\mathcal{T}_{\theta\theta}(r)$).
Third,
the last term is a contribution from the Higgs potential.
The contribution of this term is negative for 
all spatial components $\mathcal{T}_{zz}(r)$, $\mathcal{T}_{rr}(r)$,
and $\mathcal{T}_{\theta\theta}(r)$, while the contribution to
$\mathcal{T}_{00}(r)$ is positive.
These contributions are understood as the negative pressure
owing to the instability of a state having a deviation of $\phi$
from its vacuum expectation value $v$.
Fourth,
because $\mathcal{T}_{rr}(r)=\mathcal{T}_{\theta\theta}(r)$ at $r=0$,
one obtains $(\partial_r\phi)^2=\phi^2\tilde{A}^2/r^2$ in the $r\to0$ limit.
As a result,
the signs of $\mathcal{T}_{rr}(r)$ and $\mathcal{T}_{\theta\theta}(r)$ at
$r=0$ is determined by the interplay between first and fourth terms,
i.e. contributions from the gauge field and the Higgs potential,
respectively.

\section{Magnetic vortex}
\label{sec:vortex}

In this section, we discuss the magnetic monopoles and
the classical solution of the magnetic vortex between the monopoles
in the AH model with and without boundaries.

\subsection{Magnetic vortex with finite length}
\label{sec:mag_vor}

Let us first consider a magnetic vortex with the finite length $R$
between two magnetic monopoles with opposite charges.
As shown by Dirac~\cite{Dirac:1931kp}, Maxwell theory can have
monopoles with quantized charges.
These monopoles are associated with a singularity of the gauge field
called the Dirac string.
When two monopoles with opposite charges are 
at $(r,z)=(0,\pm R/2)$,
the Dirac string can be located on the $z$ axis between two monopoles.
The gauge field in this case is given by
\begin{align}
 \vec{A}^D(r,z)=-\frac{n}{2r}\biggl(\frac{z+R/2}{\sqrt{r^2+(z+R/2)^2}}
 -\frac{z-R/2}{\sqrt{r^2+(z-R/2)^2}}\biggr)\vec{e}_\theta
 =\frac{\tilde{A}^D(r,z)}{r}\vec{e}_\theta ,
 \label{eq:A_D}
\end{align}
where the winding number $n$ corresponds to 
the charge of a monopole at $z=R/2$ in the unit of $2\pi/g$.
One easily finds that 
the magnetic field $\vec{B}^D=\nabla\times \vec{A}^D$ 
is given by the superposition of the Coulombic
fields with the charges $\pm2n\pi/g$ at $(r,z)=(0,\pm R/2)$.
The magnetic field at the origin thus is given by
\begin{align}
 |\vec{B}(0)|= 2 \frac{2n\pi}g \frac1{4\pi (R/2)^2} = \frac{4n}{gR^2}.
 \label{eq:|B(0,0)|}
\end{align}
Substituting Eq.~(\ref{eq:|B(0,0)|}) into Eq.~(\ref{eq:T=diagE}),
one obtains
\begin{align}
 \mathcal{T}_{00}(0)=\mathcal{T}_{rr}(0)=\mathcal{T}_{\theta\theta}(0)
 = \frac{8n^2}{g^2R^4} ,
 \quad
 \mathcal{T}_{zz}(0) = -\frac{8n^2}{g^2R^4},
 \label{eq:coul_behave}
\end{align}
on the mid-plane.
In what follows, we consider the vortex with $n=1$.

The Dirac monopoles can also be introduced in the AH model.
As the gauge field in this case has the same singularity as
Eq.~(\ref{eq:A_D}), when monopoles are located at $(r,z)=(0,\pm R/2)$
it is convenient to denote the gauge field
as~\cite{Koma:2003gq,Maedan:1989ju,Kodama:1997zc}
\begin{align}
 &\vec{A}(r,z)=\vec{A}^D(r,z)+\vec{a}(r,z),
 \\
 &\vec{a}=a(r,z)\vec{e}_\theta
 =\frac{\Tilde{a}(r,z)}{r}\vec{e}_\theta.
\end{align}
Here, $\Tilde{a}(r,z)$ does not have a singularity and represents
the deviation of $\vec{A}(r,z)$ from Eq.~(\ref{eq:A_D}).
The classical solution of the magnetic vortex is then obtained
by minimizing the total energy
\begin{align}
 E=\int d^3x\,\mathcal{T}_{00}(r,z)
 =2\pi\int_0^\infty rdr\int_{-\infty}^\infty dz\,
 \mathcal{T}_{00}(r,z) \label{eq:energy},
\end{align}
as a functional of $\Tilde{a}(r,z)$ and $\phi(r,z)$
with the boundary conditions
\begin{align}
 &\tilde{a}(r,z) \rightarrow0
 &&\mathrm{for}\,\,r\rightarrow0,\,\,\forall  z
 \label{eq:bc1-fin_a}
 \\
 &\phi(r,z) \rightarrow0 \quad
 &&\mathrm{for}\,\,r\rightarrow0,\,\,-R/2\leq z\leq R/2, 
 \label{eq:bc1-fin_phi}
 \\
 &\tilde{a}(r,z) \rightarrow-\tilde{A}^D(r,z),\,
 \phi(r,z) \rightarrow v 
 &&\mathrm{for}\,\,r,\,z\rightarrow\infty.
 \label{eq:bc2-fin}
\end{align}
The condition Eq.~(\ref{eq:bc2-fin}) 
ensures that the total energy of this system is finite.

We note that
$\tilde{a}(r,z)$ and $\phi(r,z)$ hardly change within the length 
$\xi_A$ and $\xi_\chi$.
Therefore, when the condition $R\ll\xi_A,\xi_\chi$ is satisfied,
they remain Eqs.~(\ref{eq:bc1-fin_a}) and (\ref{eq:bc1-fin_phi})
and take values almost equal to zero
around the monopoles.
In this case, the EMT around the monopoles 
should be dominated by the contribution from 
the gauge field $\vec{A}^D(r,z)$.

\subsection{Infinitely-long vortex}
\label{sec:infinite}

Next we consider the magnetic vortex with an infinite length.
This solution is obtained by taking the $R\to\infty$ limit
in the above argument.
In this limit, the system has a translational symmetry
along $z$ direction, and $\phi$ and $\tilde{a}$ are given by
functions of only $r$ as $\phi=\phi(r)$ and $\tilde{a}=\tilde{a}(r)$.
By taking the $R\to\infty$ limit of Eq.~(\ref{eq:A_D}) 
with $n=1$ one obtains
\begin{align}
 \tilde{A}^D(r,z)=-1.
\end{align}
The vortex solution is obtained by minimizing
energy per unit length
\begin{align}
 \int_0^\infty (2\pi r) dr {\cal T}_{00}(r)
 \label{eq:tot-ene}
\end{align}
with respect to $\phi(r)$ and $\tilde{a}(r)$
with the boundary conditions 
\begin{align}
 &\tilde{a}(r) \rightarrow0,\,\phi(r)\rightarrow0
 &&\quad\mathrm{for~} r\rightarrow0, \label{eq:bc1-inf}
 \\
 &\tilde{a}(r) \rightarrow 1,\,\phi(r)\rightarrow v
 &&\quad\mathrm{for~} r\rightarrow \infty. \label{eq:bc2-inf}
\end{align}

It is convenient to introduce dimensionless variables
\begin{align}
 \rho=rgv, \quad P(\rho)=\frac{\phi(\rho/gv)}{v}, \quad
 Q(\rho)=\tilde{A}(\rho/gv).
 \label{eq:dimless_var}
\end{align}
Using these variables, the components of the dimensionless EMT
\begin{align}
 \hat{\mathcal T}_{\gamma\gamma'}(\rho)
 = \frac{\xi_A^2}{v^2} \mathcal{T}_{\gamma\gamma'}\Big(\frac{\rho}{gv}\Big)
 \label{eq:dimlessEMT}
\end{align}
are given by 
\begin{align}
 \hat{\mathcal T}_{00}(\rho) = -\hat{\mathcal T}_{zz}(\rho)
 &= \hat{T}_{(1)}(\rho) + \hat{T}_{(2)}(\rho) 
 + \hat{T}_{(3)}(\rho) + \hat{T}_{(4)}(\rho),
 \label{eq:dimlessEMT0z}
 \\
 \hat{\mathcal T}_{rr}(\rho)
 &= \hat{T}_{(1)}(\rho) + \hat{T}_{(2)}(\rho) 
 - \hat{T}_{(3)}(\rho) - \hat{T}_{(4)}(\rho),
 \label{eq:dimlessEMTr}
 \\
 \hat{\mathcal T}_{\theta\theta}(\rho)
 &= \hat{T}_{(1)}(\rho) - \hat{T}_{(2)}(\rho) 
 + \hat{T}_{(3)}(\rho) - \hat{T}_{(4)}(\rho),
 \label{eq:dimlessEMTtheta}
\end{align}
with 
\begin{align}
  \hat{T}_{(1)}(\rho) = \frac{(\partial_\rho Q)^2}{4\rho^2}  ,
  \quad
  \hat{T}_{(2)}(\rho) = \frac{(\partial_\rho P)^2}{2},
  \quad
  \hat{T}_{(3)}(\rho) = \frac{P^2Q^2}{2\rho^2},
  \quad
  \hat{T}_{(4)}(\rho) = \frac{\kappa^2(P^2-1)^2}{2}.
  \label{eq:T(i)}
\end{align}
Energy per unit length Eq.~(\ref{eq:tot-ene}) is represented as
\begin{align}
  &\int_0^\infty (2\pi r) dr {\cal T}_{00}(r)
  = 2\pi v^2 \hat\Sigma[P,Q],
\end{align}
with
\begin{align}
  &\hat\Sigma[P,Q] = 2\int_0^\infty \rho d\rho \hat{\cal T}_{00}(\rho)
  = \int_0^\infty \rho d\rho
  \biggl[\frac{1}{2\rho^2}
    (\partial_\rho Q)^2 + (\partial_\rho P)^2
    +\frac{P^2Q^2}{\rho^2}
    +\kappa^2(P^2-1)^2\biggr].
  \label{eq:hatsigma}
\end{align}
In Eq.~(\ref{eq:hatsigma}), the form of $\hat\Sigma[P,Q]$ 
depends on parameters in the AH model only through $\kappa$.
Therefore, $P(\rho)$ and $Q(\rho)$ of the vortex solution,
and accordingly the dimensionless EMT Eq.~(\ref{eq:dimlessEMT}),
depend only on $\kappa$.
Also, the energy per unit length, i.e. the string tension
of the vortex, $\sigma_{\rm AH}$, is given by 
\begin{align}
  \sigma_{\rm AH} = 2\pi v^2 \hat\sigma_{\rm AH}(\kappa),
  \label{eq:sigma_AH}
\end{align}
where $\hat\sigma_{\rm AH}(\kappa)$ is obtained by substituting
the vortex solution into $\hat\Sigma[P,Q]$.
As shown in Appendix~\ref{sec:analytic}, 
it is possible to show $\hat\sigma_{\rm AH}(1/\sqrt{2})=1$
analytically~\cite{Bogomolnyi:1976}.

\subsection{Physical units}
\label{sec:units}

\begin{figure}
 \centering
 \includegraphics[width=10cm,clip]{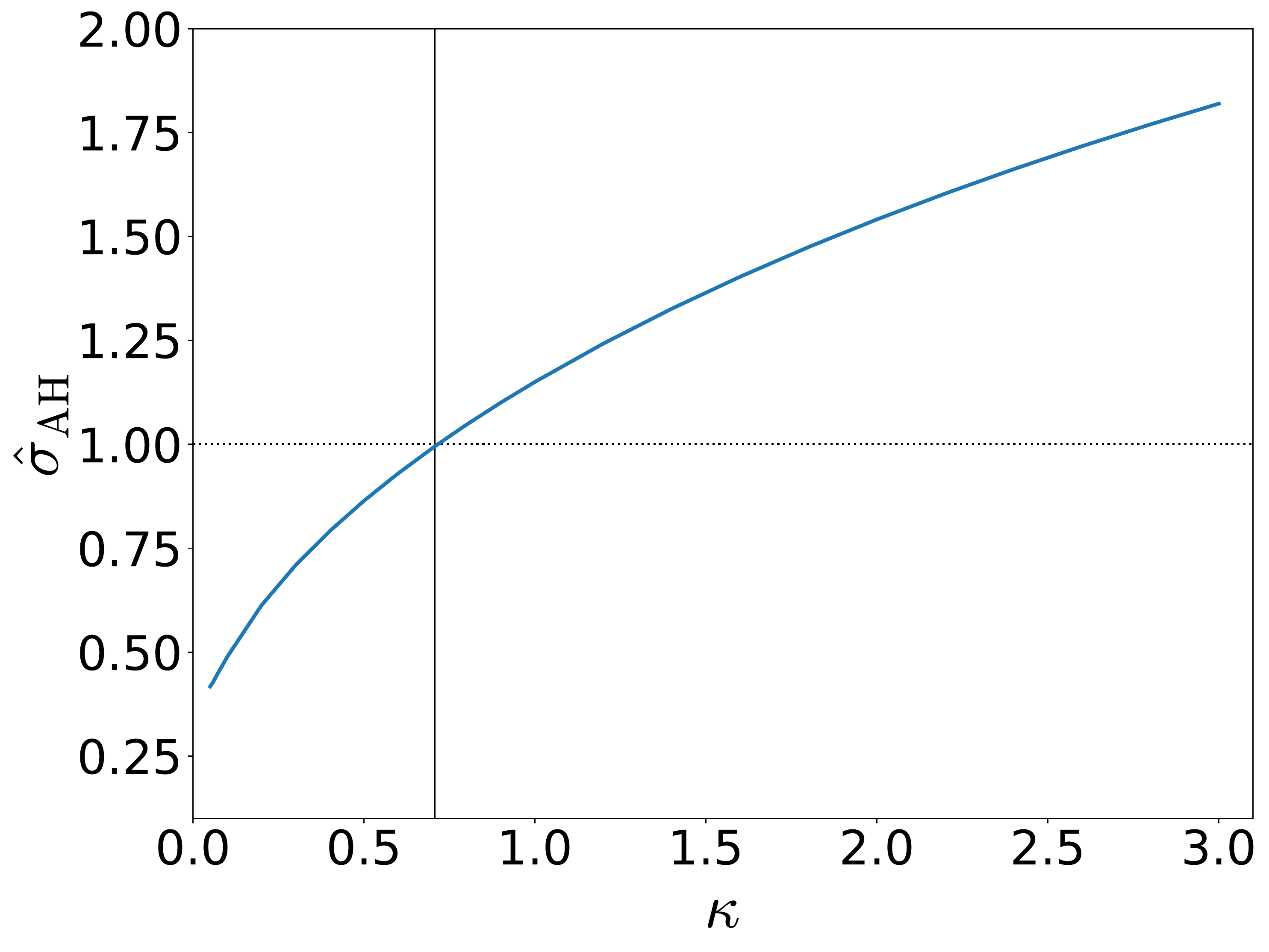}
 \caption{
 Behavior of $\hat\sigma_{\rm AH}(\kappa)$.
 The vertical line shows the Bogomol'nyi bound $\kappa=1/\sqrt{2}$.
 }
 \label{fig:k_vs_sigma}
\end{figure}

To compare the stress-tensor distribution around the magnetic vortex
obtained in the AH model with the flux tube 
in Ref.~\cite{Yanagihara:2018qqg},
it is desirable to introduce the physical dimension to the former.
The only parameter having a mass dimension in the AH model is $v$.
To determine this quantity in physical units, 
we require that the string tension of the magnetic vortex
$\sigma_{\rm AH}$ is equivalent with the string tension of the flux tube
in SU(3) YM theory, $\sigma_{\rm YM}$~\cite{Suganuma:1993ps,Kodama:1997zc}.
From Eq.~(\ref{eq:sigma_AH}), we then have
\begin{align}
  v = \sqrt{ \frac{\sigma_{\rm YM}}{2\pi \hat\sigma_{\rm AH}(\kappa)} }.
  \label{eq:v}
\end{align}
The value of $\hat\sigma_{\rm AH}(\kappa)$ is obtained numerically
from the solution of the magnetic vortex with an infinite length.
In Fig.~\ref{fig:k_vs_sigma}, we show the behavior of
$\hat\sigma_{\rm AH}(\kappa)$ as a function of $\kappa$.
The figure shows that $\hat{\sigma}_{\rm AH}=1$ at $\kappa=1/\sqrt2$.
This property can be shown analytically~\cite{Bogomolnyi:1976},
as discussed in Appendix~\ref{sec:analytic}.

For the value of $\sigma_{\rm YM}$, we use
\begin{align}
  \sigma_{\rm YM} = 1.132(10)~{\rm GeV/fm},
  \label{eq:sigma_YM}
\end{align}
which is obtained from the large $R$ behavior of the $Q\bar{Q}$
potential at $\beta=6.6$ in Ref.~\cite{Yanagihara:2018qqg}.

Because the right-hand side of Eq.~(\ref{eq:v}) depends only on $\kappa$,
even after fixing the value of $v$ in physical units,
there exists an arbitrariness to vary $g$ and $\lambda$ with fixed
$\kappa=\sqrt{\lambda}/g$.
This means that $\xi_\chi$ and $\xi_A$ are not determined by fixing 
$v$ and $\kappa$.
This arbitrariness is canceled out in Eq.~(\ref{eq:dimlessEMT})
for the infinitely-long case, but has to be taken into account
explicitly when $R$ is finite.

\subsection{Details of numerical analysis}

The classical solution of the magnetic vortex is obtained numerically
by minimizing the total energy with the boundary conditions
Eq.~(\ref{eq:bc1-inf}) and (\ref{eq:bc2-inf}).
For this procedure with finite length $R$
we discretize the half-plane of $r$ and $z$ and 
iteratively update the fields $\phi(r,z)$ and $\tilde{a}(r,z)$
at even and odd sites via the over-relaxation method.
We take the mesh size $\Delta$ of the lattice as
\begin{align}
  \Delta = p ~{\rm min}( \xi_A , \xi_\chi )
\end{align}
where $p$ is chosen in the range $0.02<p<0.1$ so that
the ratio of $R$ and $2\Delta$ is given by an integer.
We have checked that the mesh size with $p=0.1$ is small enough
to suppress the discretization error in the range $0.1<\kappa<1.0$
by changing the mesh size.
The spatial lengths along $r$ and $z$ directions are chosen
so that the boundaries are at least $3.4$ times 
longer than ${\rm max}(\xi_A , \xi_\chi)$.
We have checked that the finite size effects are well suppressed
with this setting.
The iteration is terminated when the 
total energy becomes unchanged in each step.
The criterion is set to be 
$\delta E=(E_{n+1}-E_n)/E_n<1.0\times10^{-6}$ for the $n$-th iteration.
Once we obtain the solution for $\phi$ and $\tilde{a}$,
EMT is obtained by substituting them
into Eqs.~(\ref{eq:emt-0})--(\ref{eq:emt-z}).

For the infinitely-long case we proceed similar procedures in the 
one dimensional space of $\rho$ for the dimensionless
functions $P(\rho)$ and $Q(\rho)$.

We note that the number of lattice points increases
when the difference between $\xi_A$ and $\xi_\chi$
is large in order to satisfy the above conditions for the mesh size
$\Delta$ and the size of the lattice.
This means that the numerical cost increases 
at small and large $\kappa$.
Because of this difficulty, we limit our numerical analysis in the
range $0.1\le\kappa\le1.0$ and $0.05\le\kappa\le3.0$ for
the finite-length and infinitely-long vortices, respectively.

\section{Numerical results}
\label{sec:result}

In this section we discuss the numerical results on the
EMT distribution around the magnetic vortex in the AH model.

\subsection{Infinitely-long vortex}

\begin{figure}
  \includegraphics[width=0.47\textwidth]{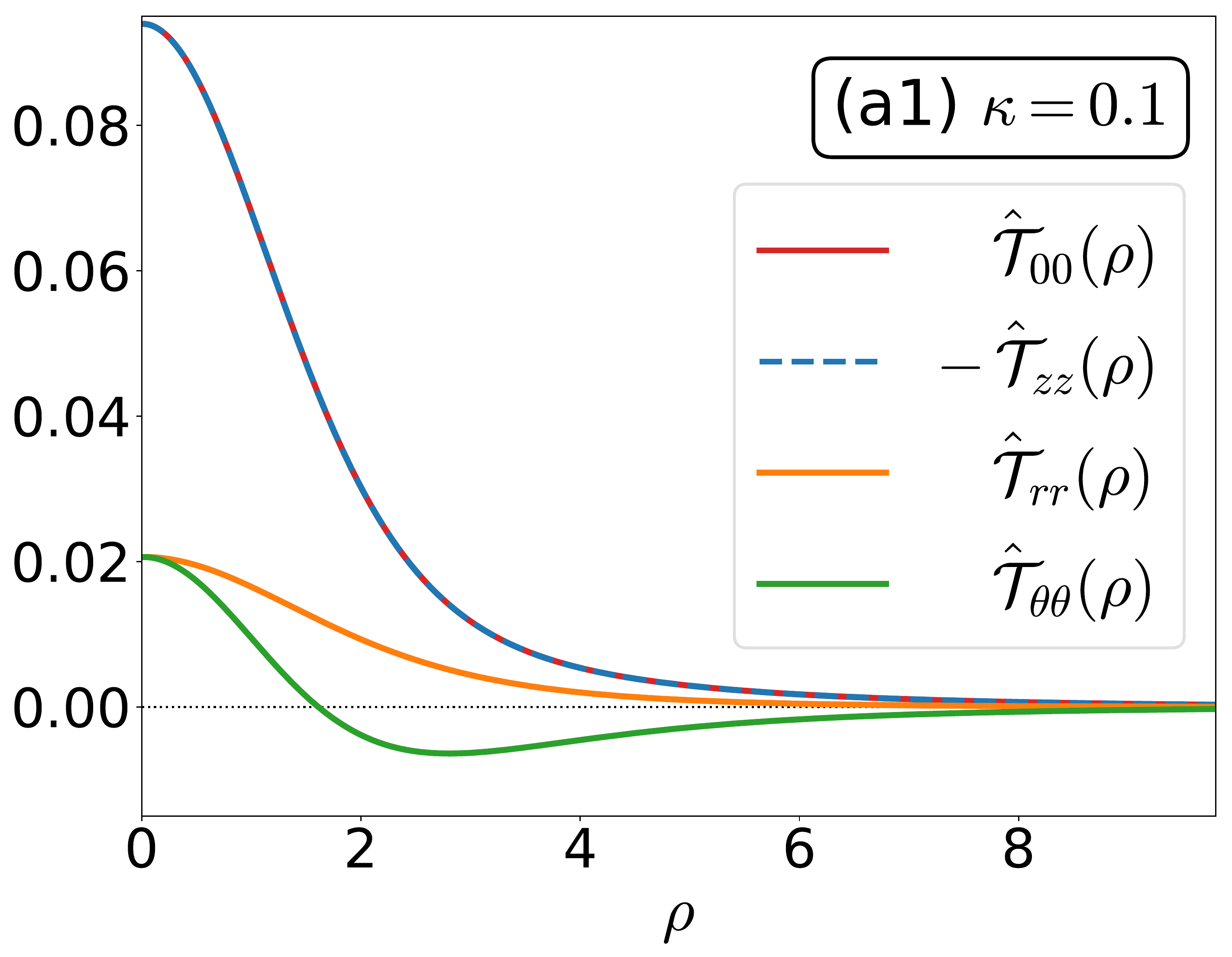}
  \includegraphics[width=0.47\textwidth]{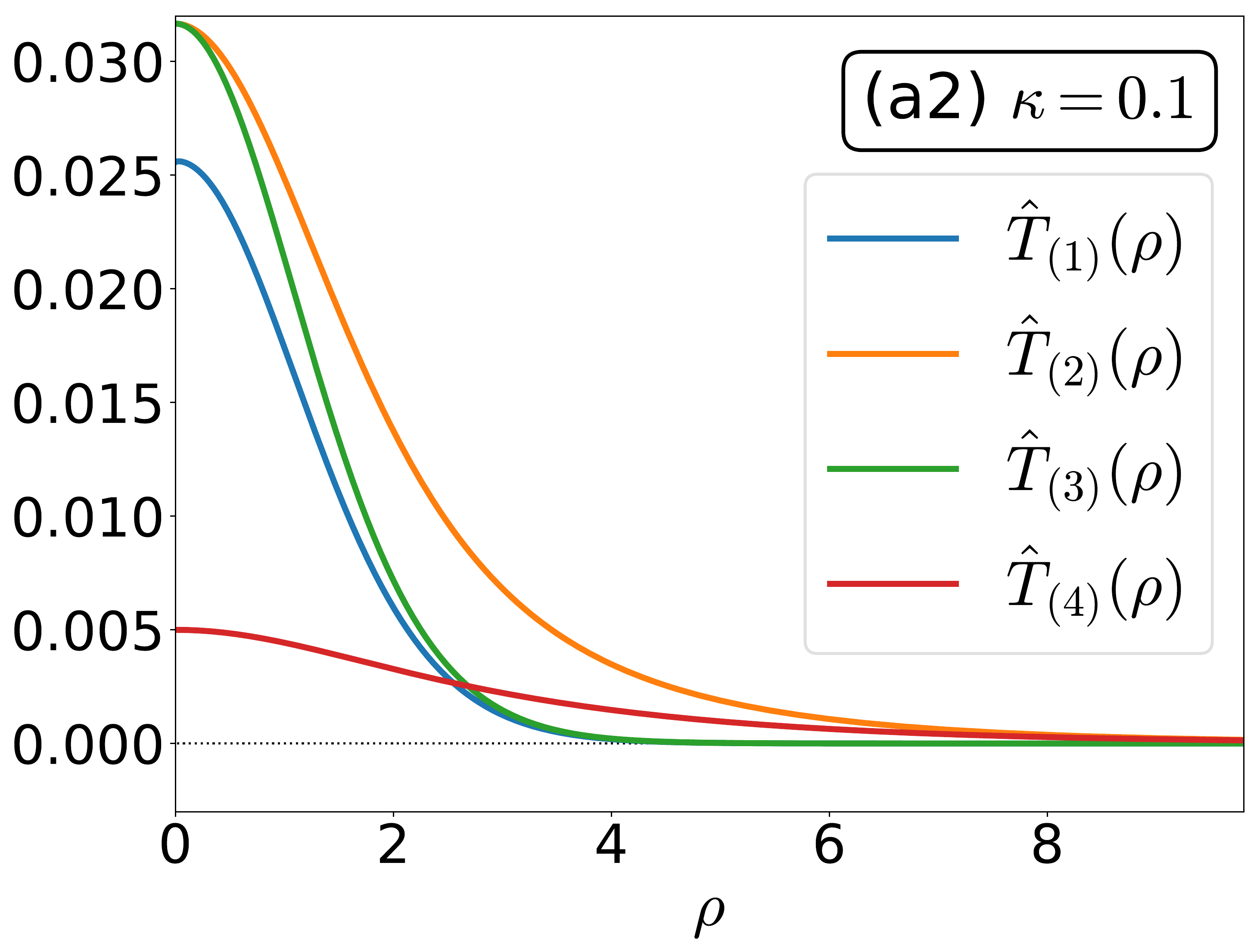}
  \\
  \includegraphics[width=0.47\textwidth]{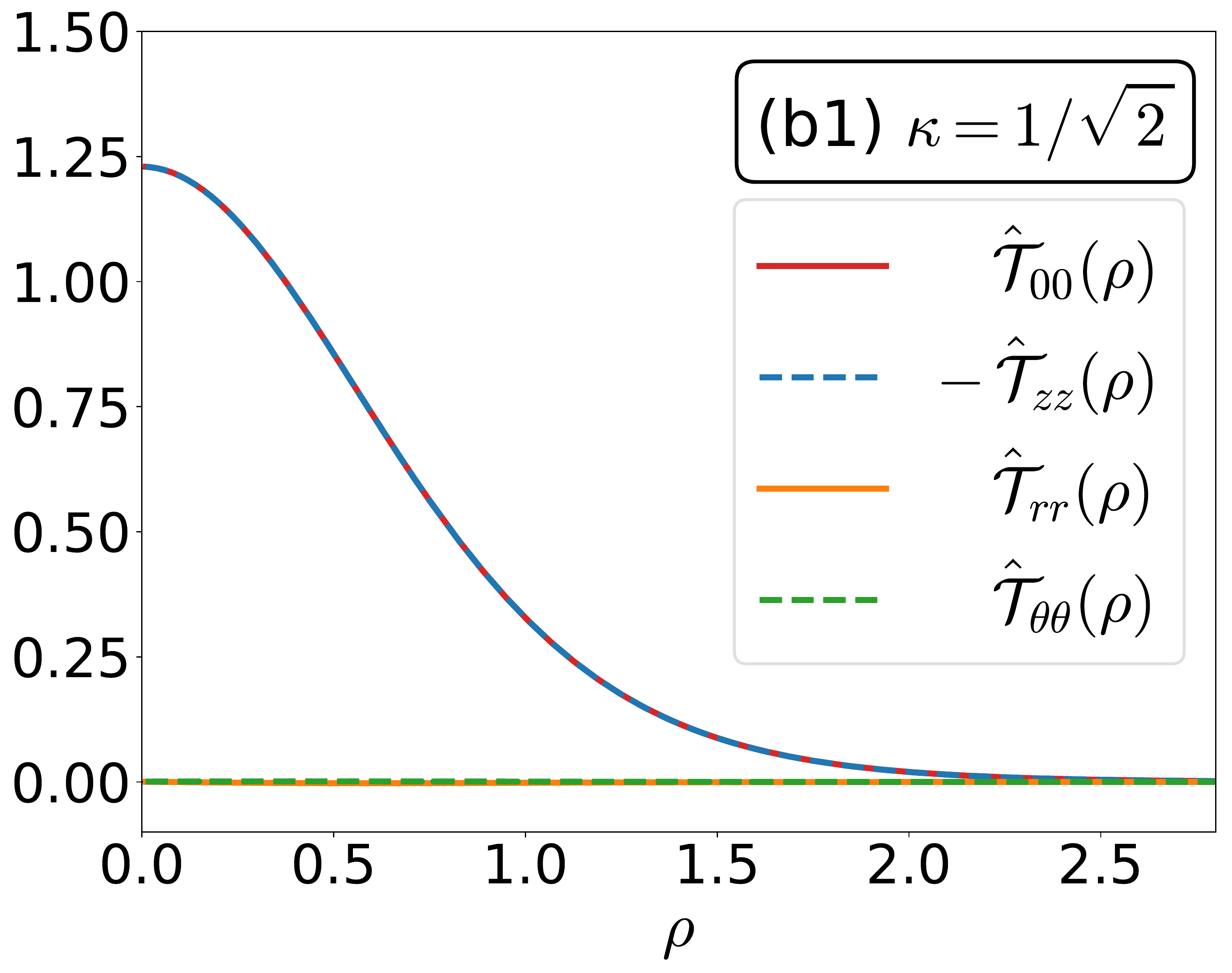}
  \includegraphics[width=0.47\textwidth]{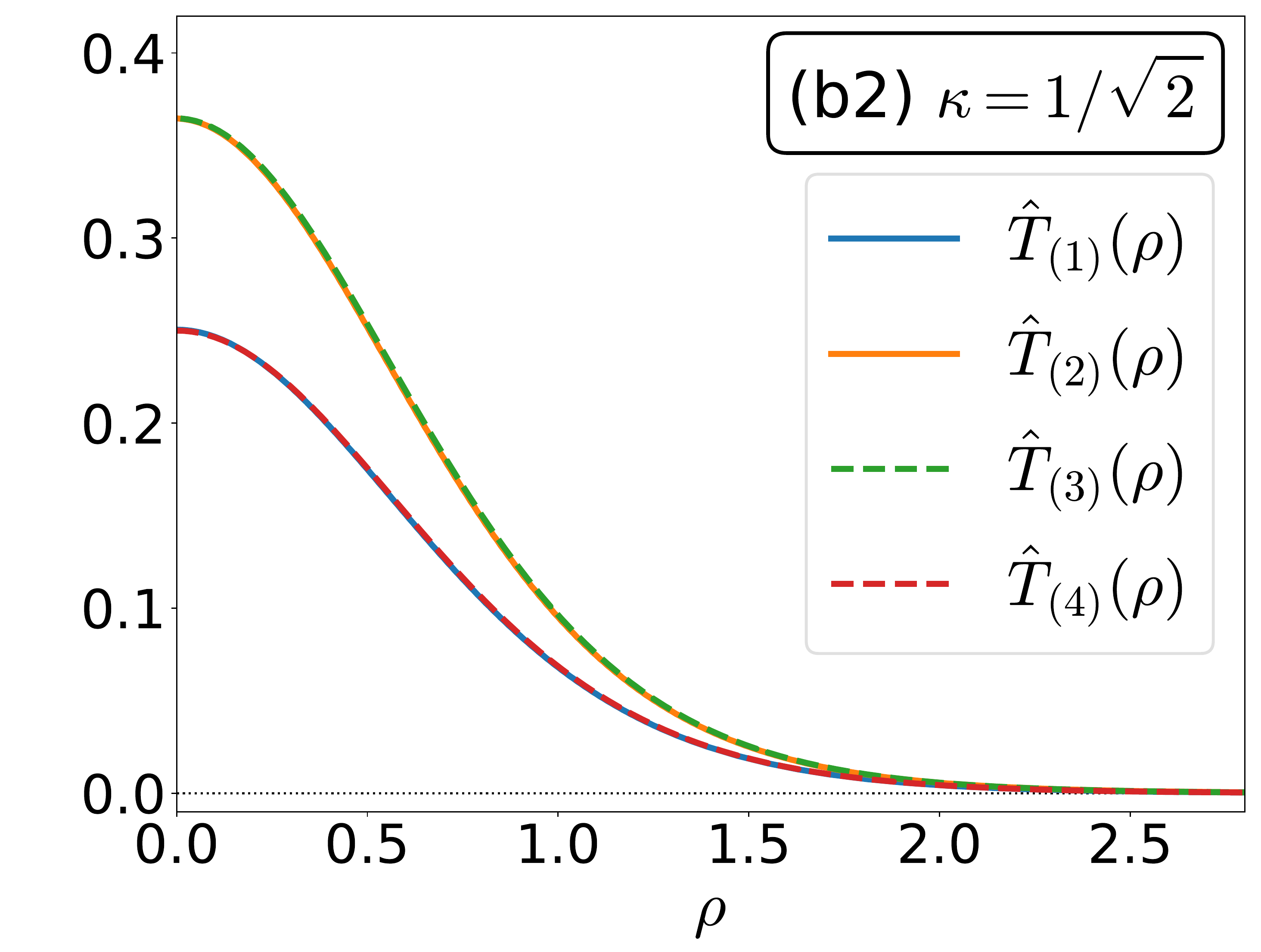}
  \\
  \includegraphics[width=0.47\textwidth]{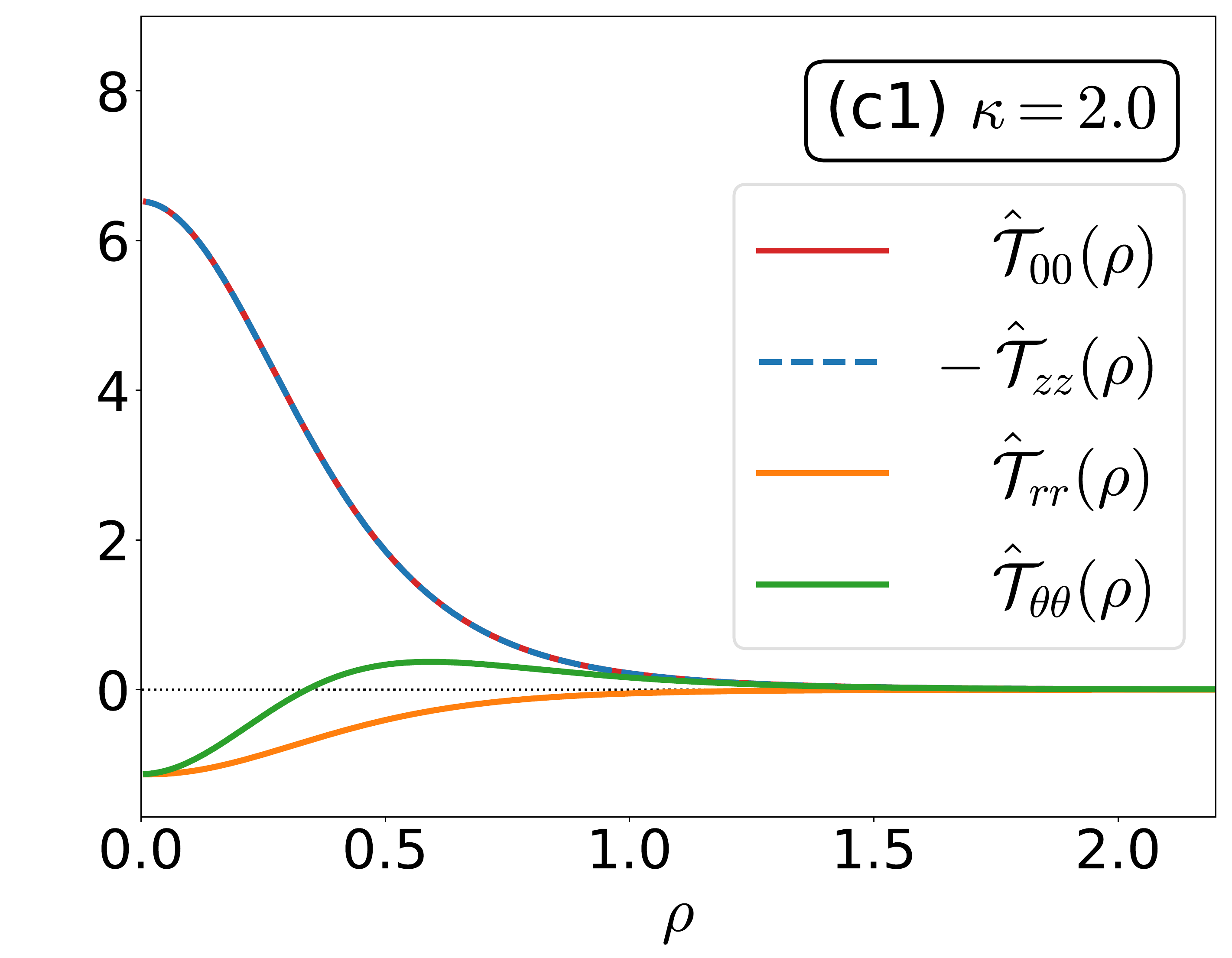}
  \includegraphics[width=0.47\textwidth]{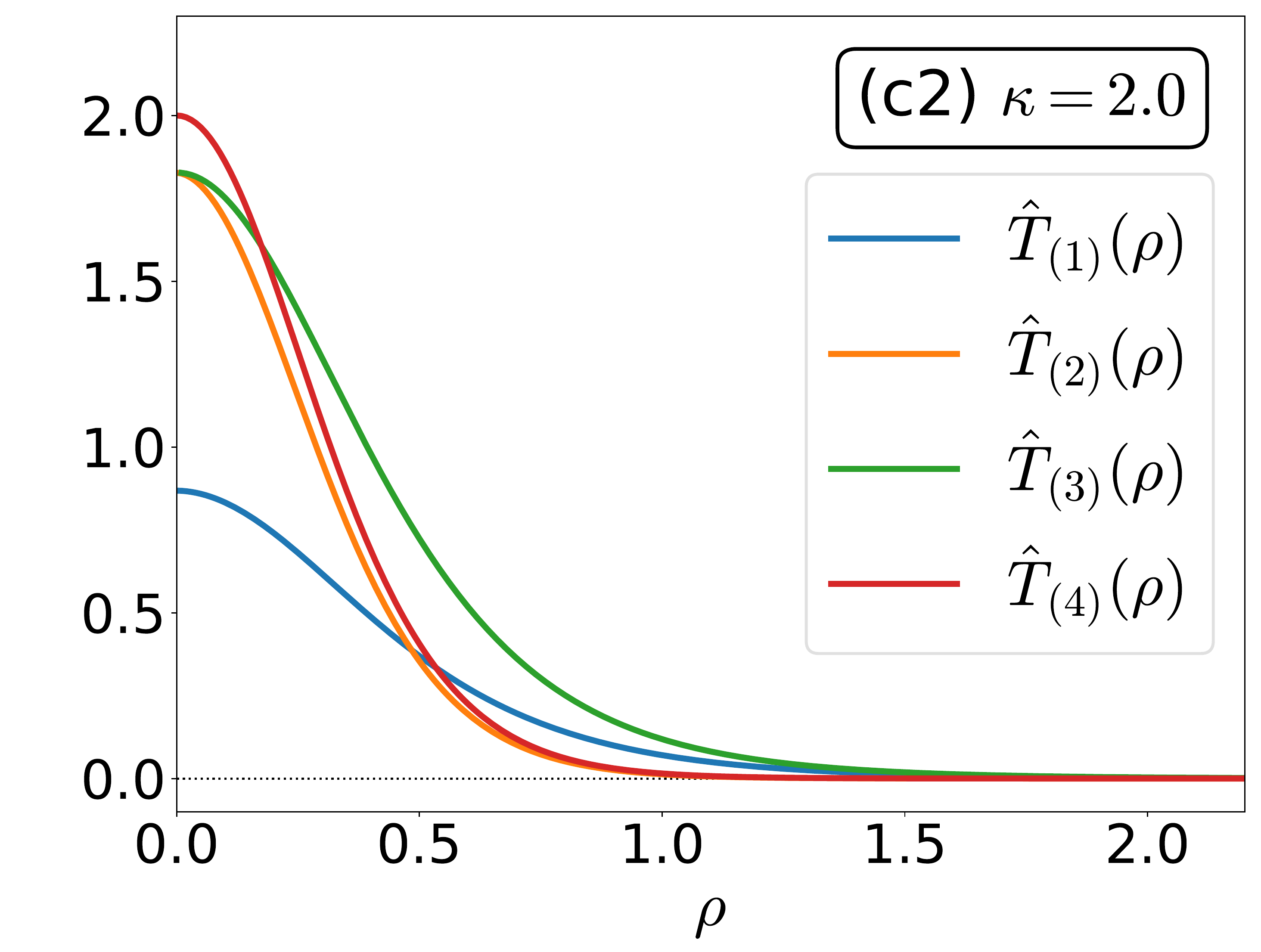}
 \caption{
   EMT distribution around
   the infinitely-long flux tube (left) and the contributions
   of individual terms in Eq.~(\ref{eq:T(i)}) (right).
   Top, middle, and bottom panels show the results for 
   $\kappa=0.1$, $\kappa=1/\sqrt{2}$, and $\kappa=2.0$,
   respectively.
 }
 \label{fig:inf}
\end{figure}

In this subsection, we first focus on the infinitely-long vortex.
In this case it is convenient to employ the dimensionless variables,
$\rho$ and $\hat{\mathcal T}_{\gamma\gamma}(\rho)$, 
introduced in Eqs.~(\ref{eq:dimless_var}) and (\ref{eq:dimlessEMT}),
which depend only on $\kappa$.
In the left panels of Fig.~\ref{fig:inf}, we show the $\rho$ dependences
of $\hat{\mathcal T}_{00}(\rho)$, $\hat{\mathcal T}_{zz}(\rho)$,
$\hat{\mathcal T}_{rr}(\rho)$, and $\hat{\mathcal T}_{\theta\theta}(\rho)$
for three values of $\kappa$;
top and lower panels show the results for 
type-I ($\kappa=0.1$) and type-II ($\kappa=2.0$) cases, respectively,
while the middle panel corresponds to
the Bogomol'nyi bound at $\kappa=1/\sqrt2$.
One finds that 
the sign of $\mathcal{T}_{rr}(r)$ is positive (negative)
at $\kappa=0.1$ ($\kappa=2.0$), while 
the middle panel shows that 
$\mathcal{T}_{rr}(r)=\mathcal{T}_{\theta\theta}(r)=0$ at $\kappa=1/\sqrt2$.
The latter property is obtained analytically
as discussed in Refs.~\cite{Bogomolnyi:1976,deVega:1976xbp} and 
summarized in Appendix~\ref{sec:analytic}.
For $\kappa\ne1/\sqrt2$, 
$\mathcal{T}_{\theta\theta}(r)$ behaves differently from
$\mathcal{T}_{rr}(r)$, and changes the sign at nonzero $r$.
This result is consistent with the discussion
based on the momentum conservation in Sec.~\ref{sec:stress}.

Now, let us compare the result in Fig.~\ref{fig:inf}
with the EMT distribution around
the flux tube in SU(3) YM theory~\cite{Yanagihara:2018qqg} shown in
Fig.~\ref{fig:mid_3lat}.
As in Fig.~\ref{fig:mid_3lat}, in SU(3) YM theory
$\mathcal{T}_{rr}(r)$ and $\mathcal{T}_{\theta\theta}(r)$ are
degenerated within statistics even at the largest $Q\bar{Q}$ distance,
$R=0.92$~fm.
The lattice result also suggests that $\mathcal{T}_{\theta\theta}(r)$
is always positive.
These results clearly contradict with Fig.~\ref{fig:inf}.
This result thus shows that the structure of the flux tube
in SU(3) YM theory with $R\le0.92$~fm cannot be understood by the
comparison with the magnetic vortex with an infinite length in the AH model.

\begin{figure}
 \centering
 \includegraphics[width=10cm,clip]{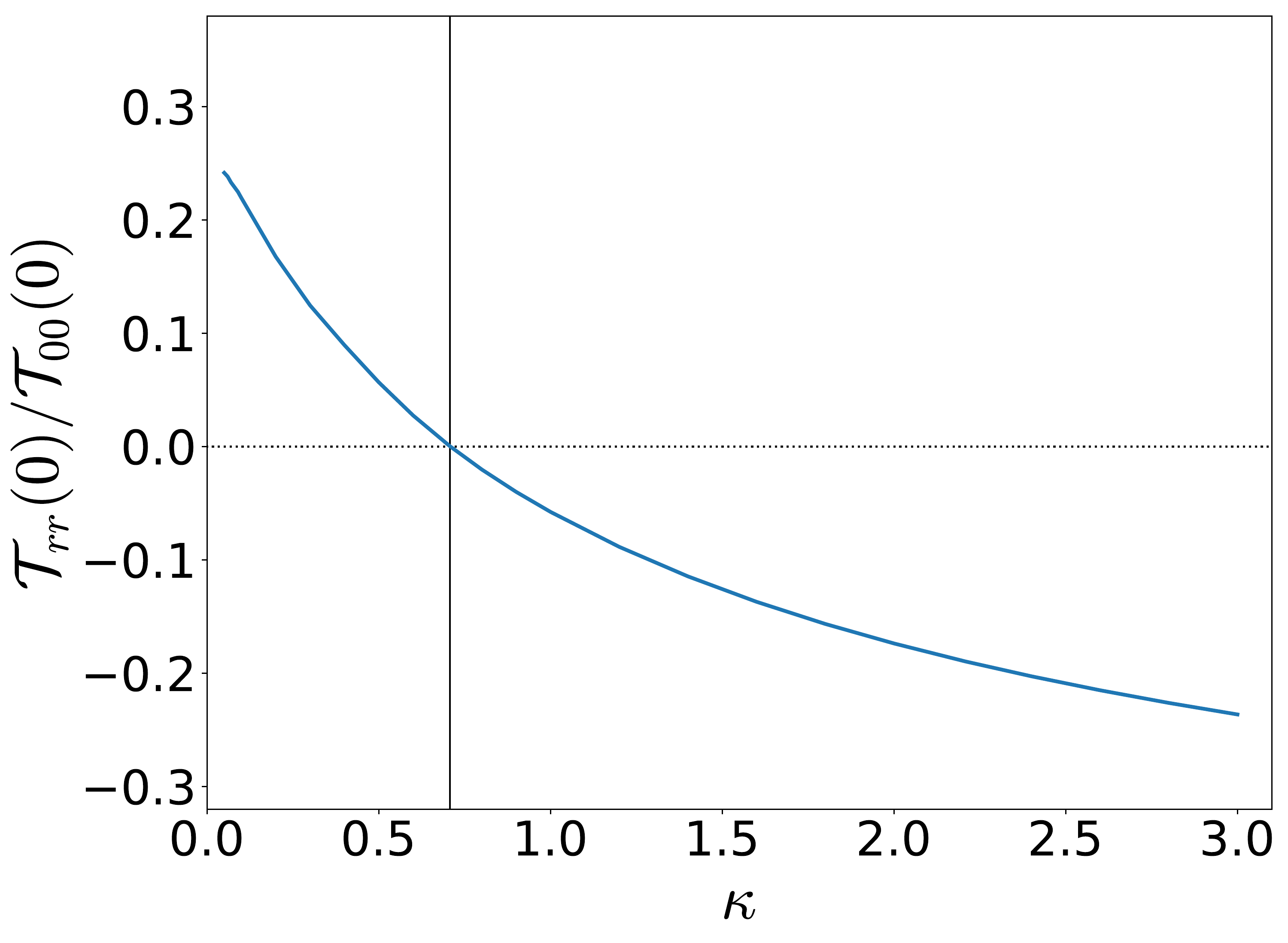}
 \caption{
 $\kappa$-dependence of the ratio
   $\mathcal{T}_{rr}(0)/\mathcal{T}_{00}(0)$ for the infinitely-long
   vortex.
   The value $\kappa=1/\sqrt{2}$ is indicated by the vertical line.
 }
 \label{fig:k_vs_T}
\end{figure}

In spite of this conclusion, 
it is instructive to take a closer look at the EMT distribution
in the AH model in Fig.~\ref{fig:inf}.
As in Eqs.~(\ref{eq:dimlessEMT0z})--(\ref{eq:dimlessEMTtheta}),
the EMT around the infinitely-long vortex consists of 
four terms in Eq.~(\ref{eq:T(i)}).
In the right panels of Fig.~\ref{fig:inf}, the behavior of
these terms are shown separately for each $\kappa$.
At $\rho=0$, one has $\hat{T}_{(2)}(\rho)=\hat{T}_{(3)}(\rho)$ which
leads to $\mathcal{T}_{rr}(0)=\mathcal{T}_{\theta\theta}(0)$.
The sign of $\mathcal{T}_{rr}(0)$ thus is determined by the interplay
between $\hat{T}_{(1)}(\rho)$ and $\hat{T}_{(4)}(\rho)$.
As discussed in Sec.~\ref{sec:EMT:AH}, 
$\hat{T}_{(1)}(\rho)$ is the Maxwell stress having a positive contribution
to $\mathcal{T}_{rr}(r)$ as in Eq.~(\ref{eq:dimlessEMTr}),
while the contribution of $\hat{T}_{(4)}(\rho)$ is negative in this channel.
A positive $\mathcal{T}_{rr}(0)$ at $\kappa<1/\sqrt{2}$ thus
means $\hat{T}_{(1)}(0) > \hat{T}_{(4)}(0)$, i.e.,
the contribution of the gauge field plays a dominant role
at the core of the vortex in the type-I region.
On the other hand, for $\kappa>1/\sqrt2$ the effect of the Higgs
potential dominates over the gauge field.

In this way, the sign of $\mathcal{T}_{rr}(0)$ can be used to distinguish
the type-I and type-II provided that the length of the vortex is infinite.
In Fig.~\ref{fig:k_vs_T},
we show the ratio $\mathcal{T}_{rr}(0)/\mathcal{T}_{00}(0)$ 
as a function of $\kappa$.
The figure shows that $\mathcal{T}_{rr}(0)/\mathcal{T}_{00}(0)$ is
a monotonic function of $\kappa$ changing the sign at $\kappa=1/\sqrt2$.

\subsection{Finite-Length Flux Tube}

Next, we investigate the magnetic vortex with finite length $R$.
In the following, numerical results are shown in physical units
by fixing the value of $v$ through Eq.~(\ref{eq:v}) in order to
make the comparison with Ref.~\cite{Yanagihara:2018qqg} easy.
After fixing the value of $v$ in physical units, 
the vortex solution with length $R$ depends 
on two parameters in the AH model.
In the following, we use $\kappa$ and $g$ as the parameters.

\begin{figure}
 \centering
 \includegraphics[width=0.325\textwidth]{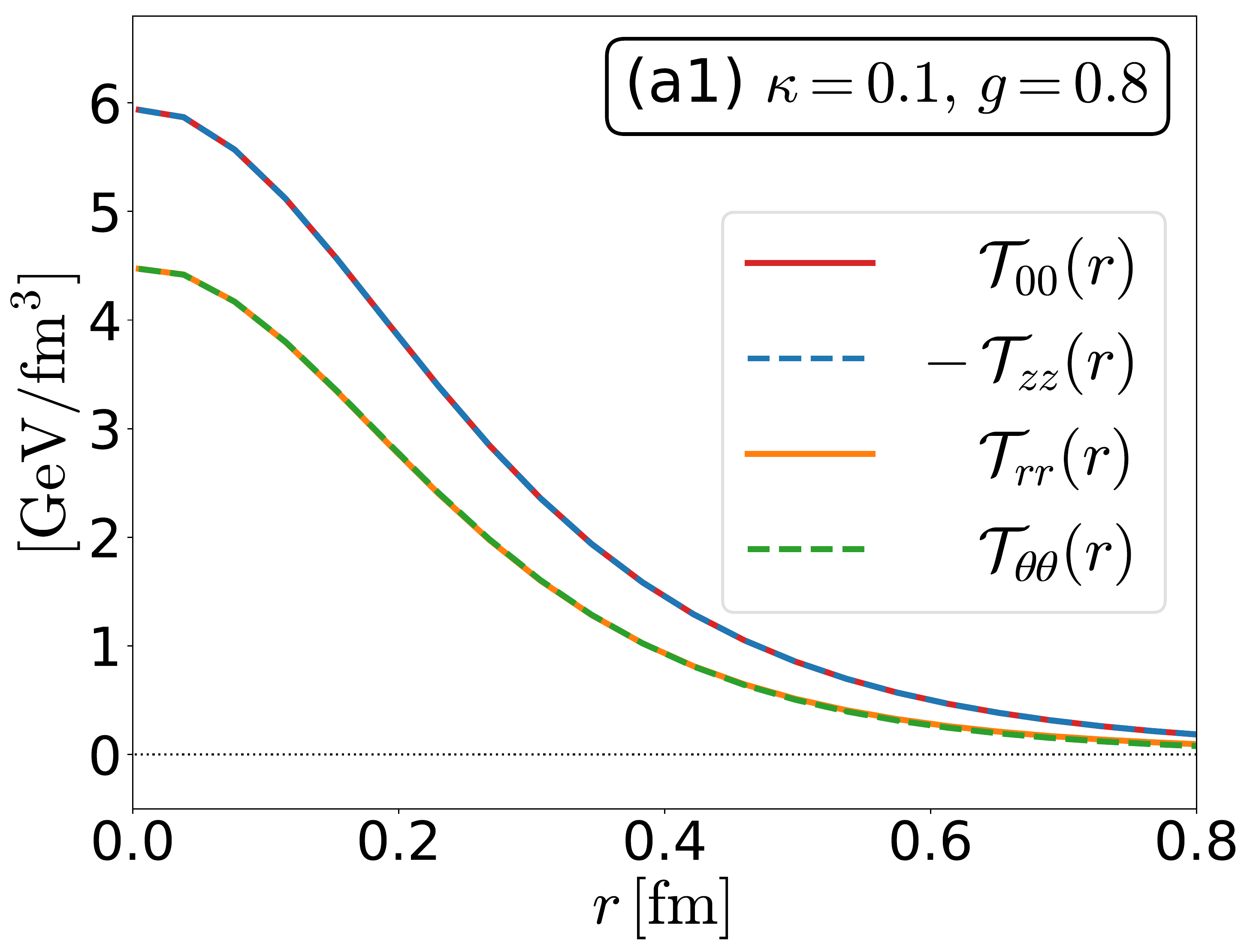}
 \includegraphics[width=0.325\textwidth]{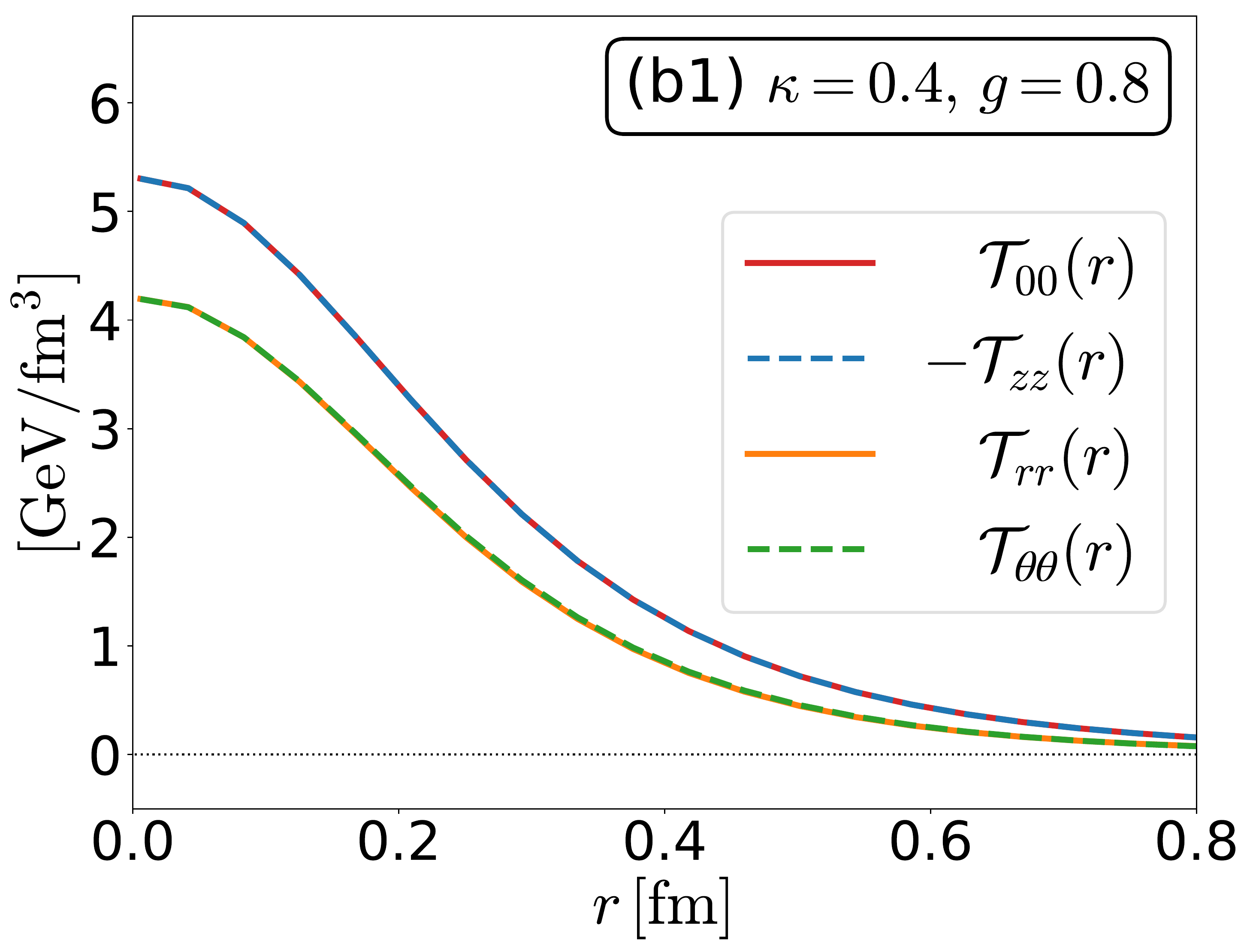}
 \includegraphics[width=0.325\textwidth]{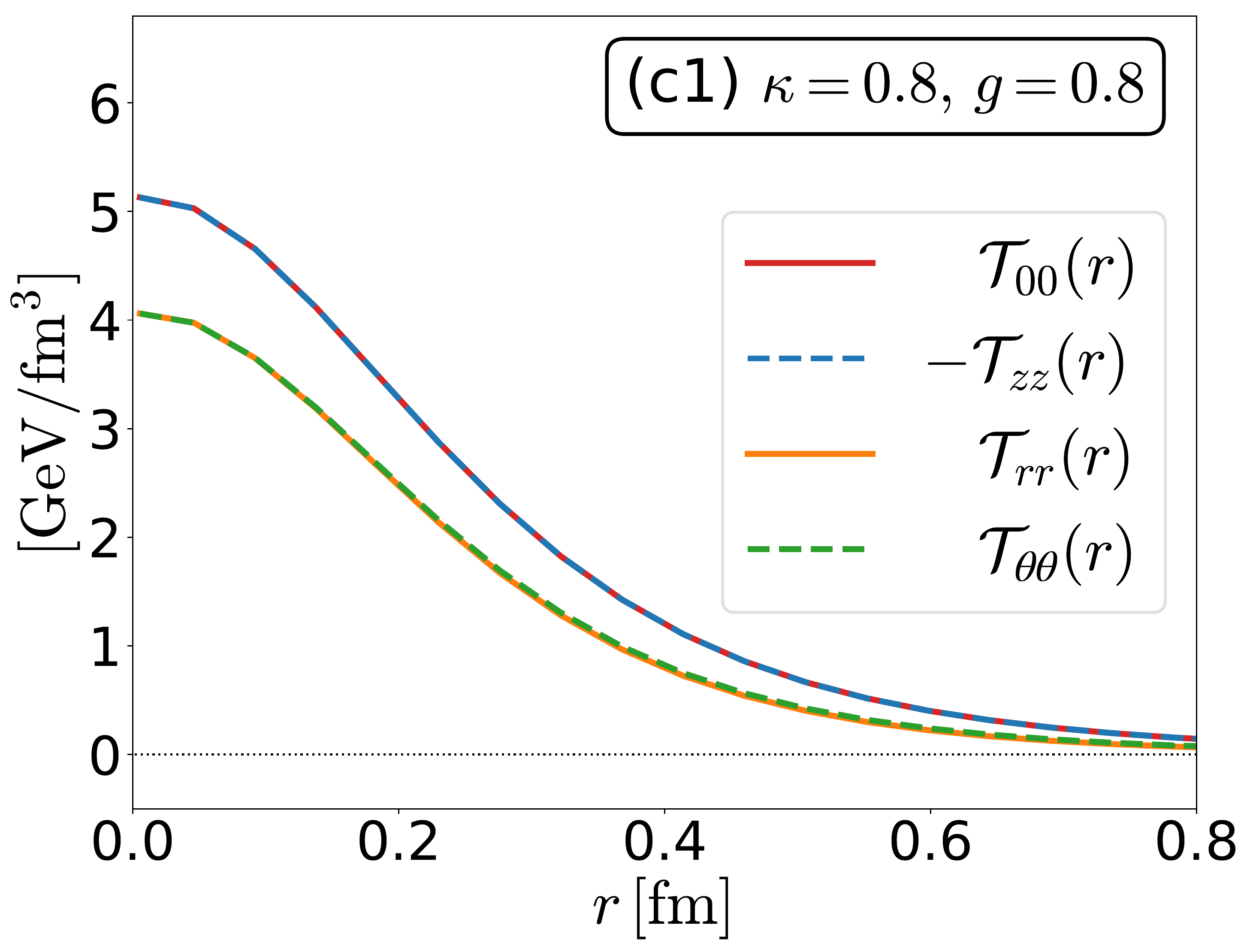}
 \\
 \includegraphics[width=0.325\textwidth]{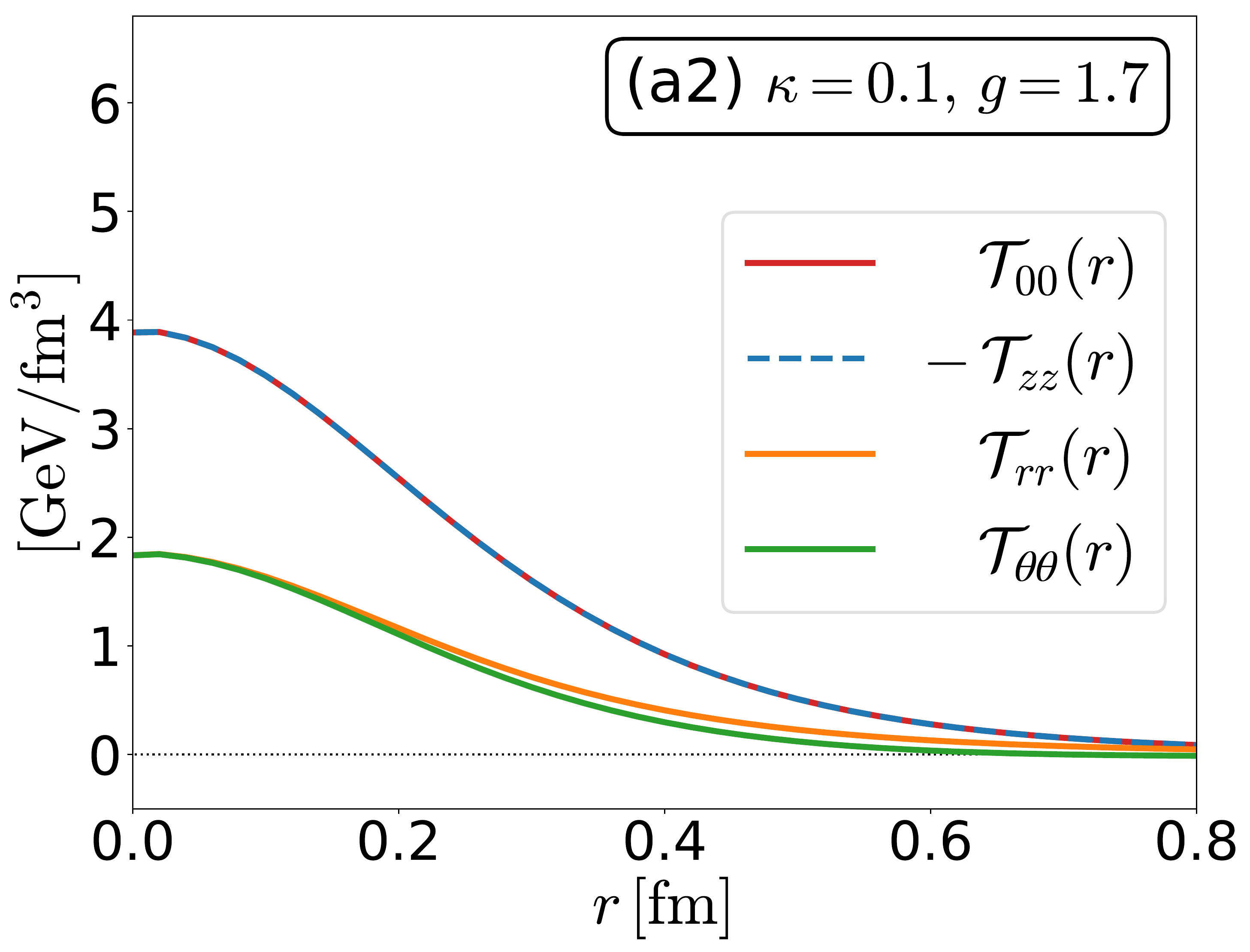}
 \includegraphics[width=0.325\textwidth]{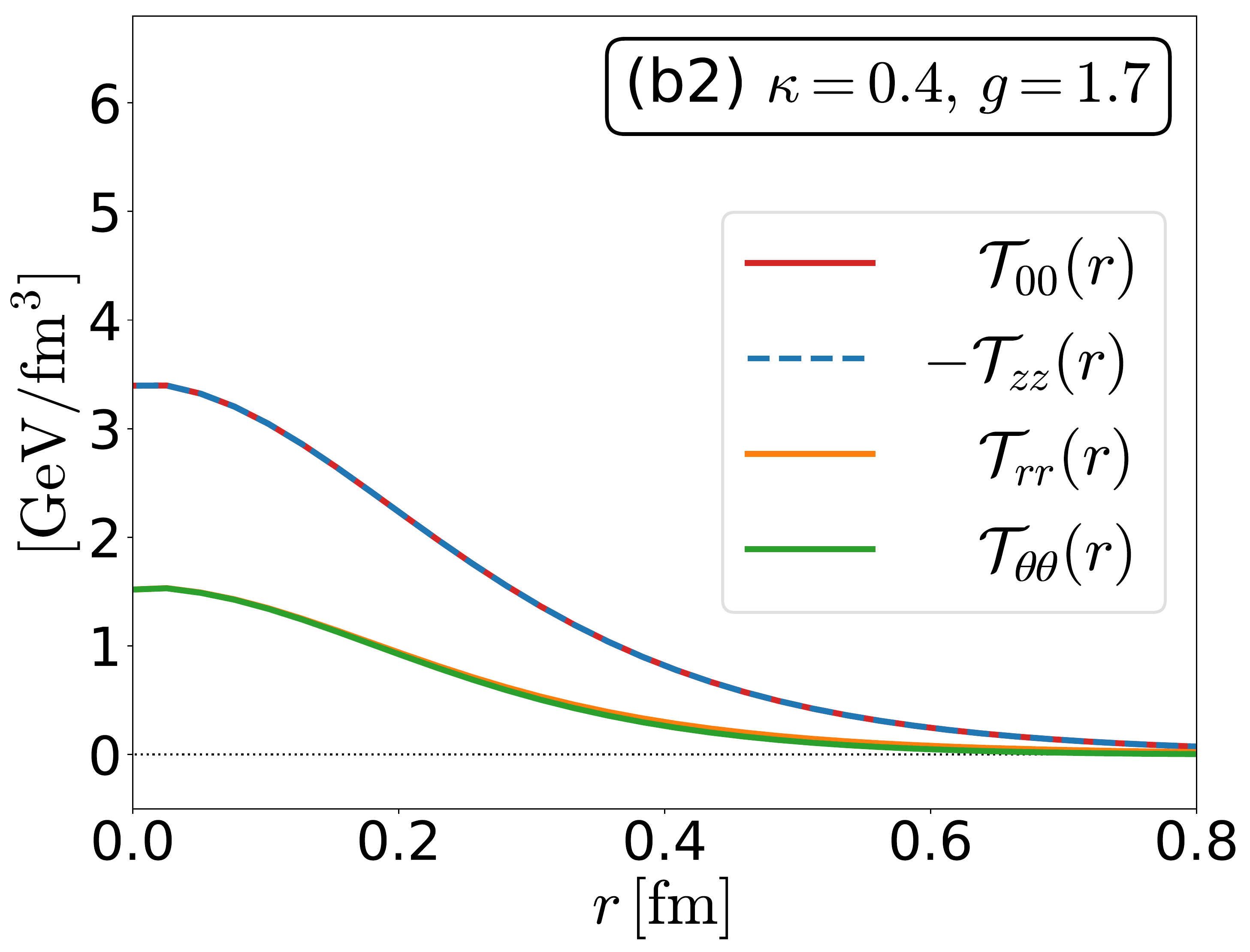}
 \includegraphics[width=0.325\textwidth]{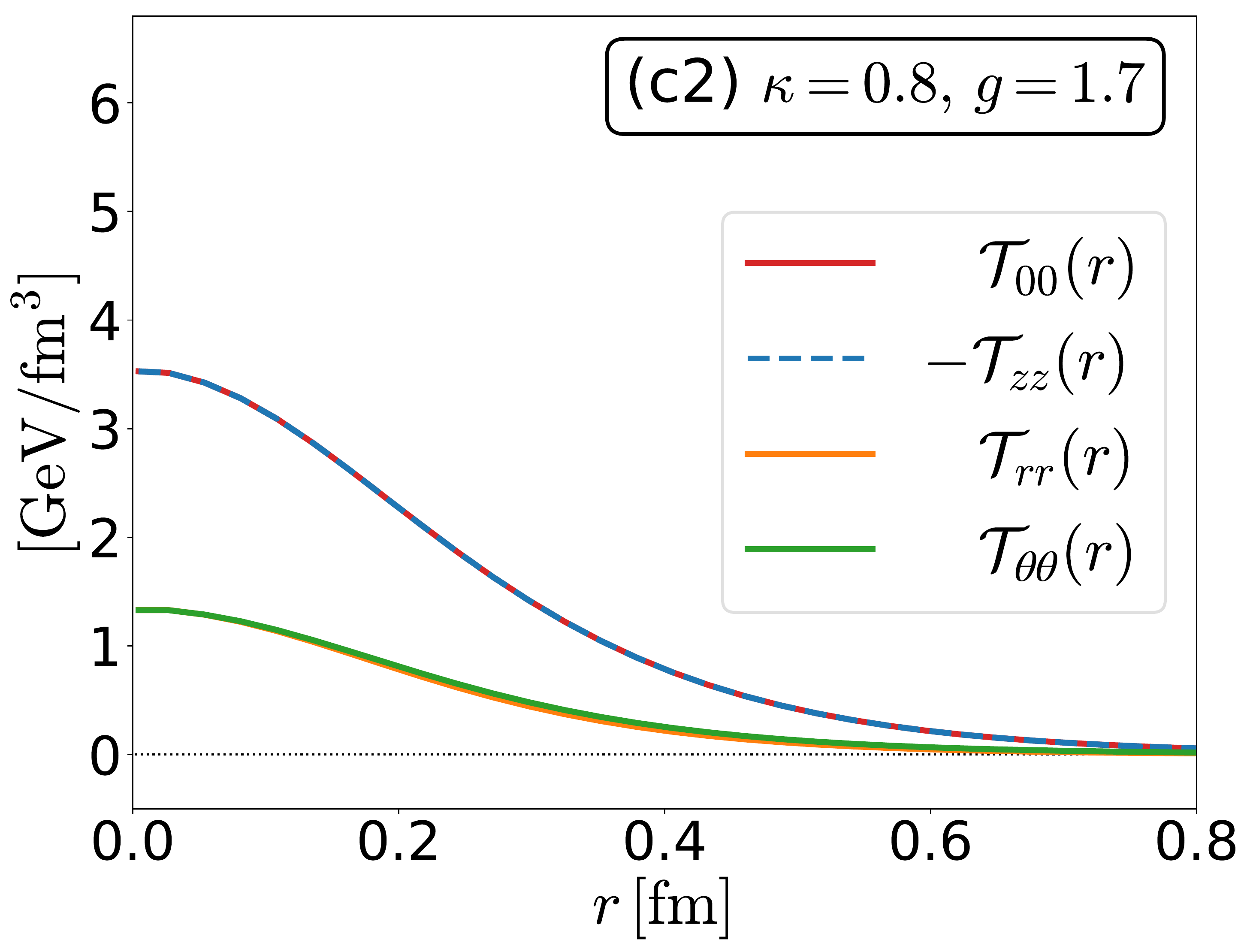}
 \\
 \includegraphics[width=0.325\textwidth]{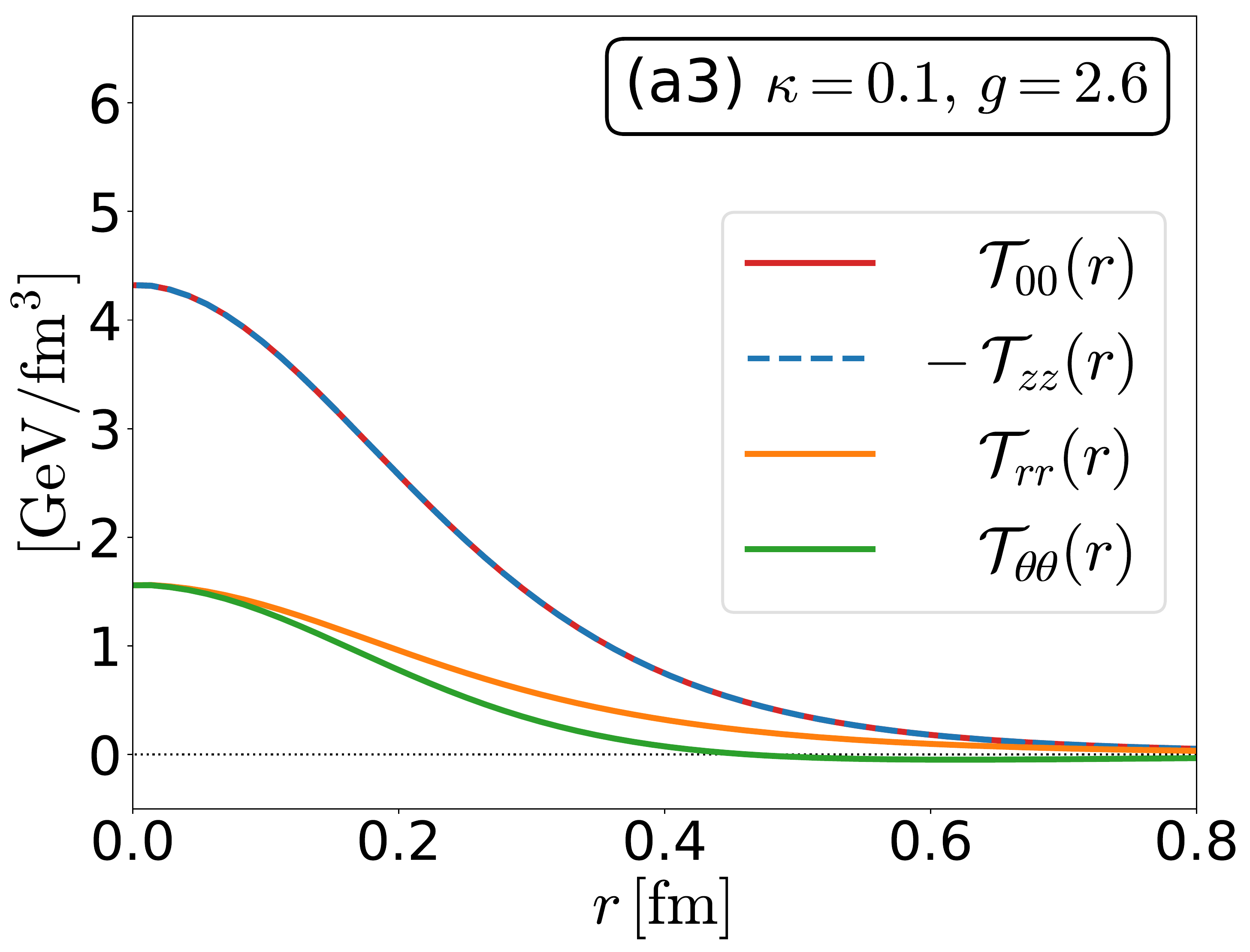}
 \includegraphics[width=0.325\textwidth]{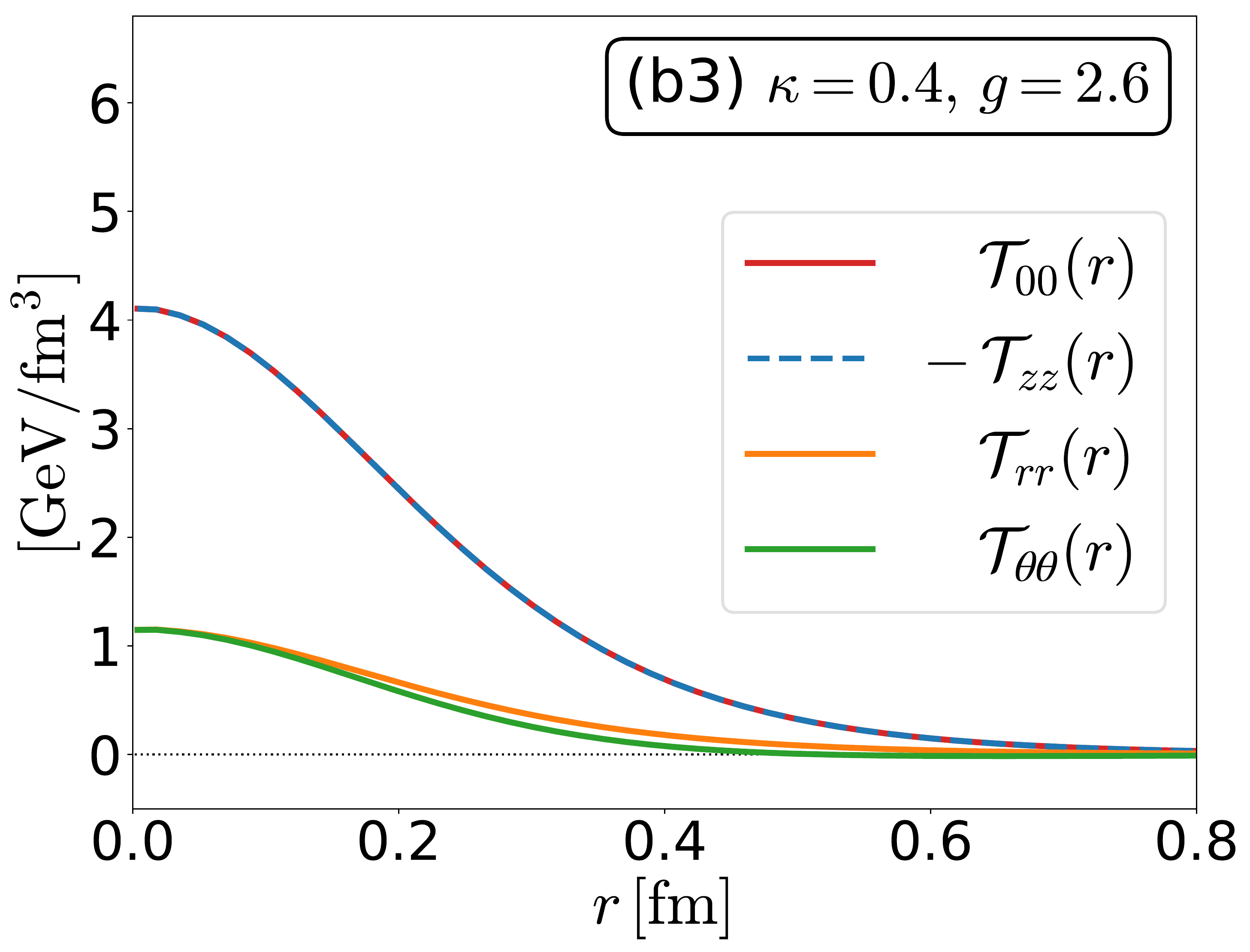}
 \includegraphics[width=0.325\textwidth]{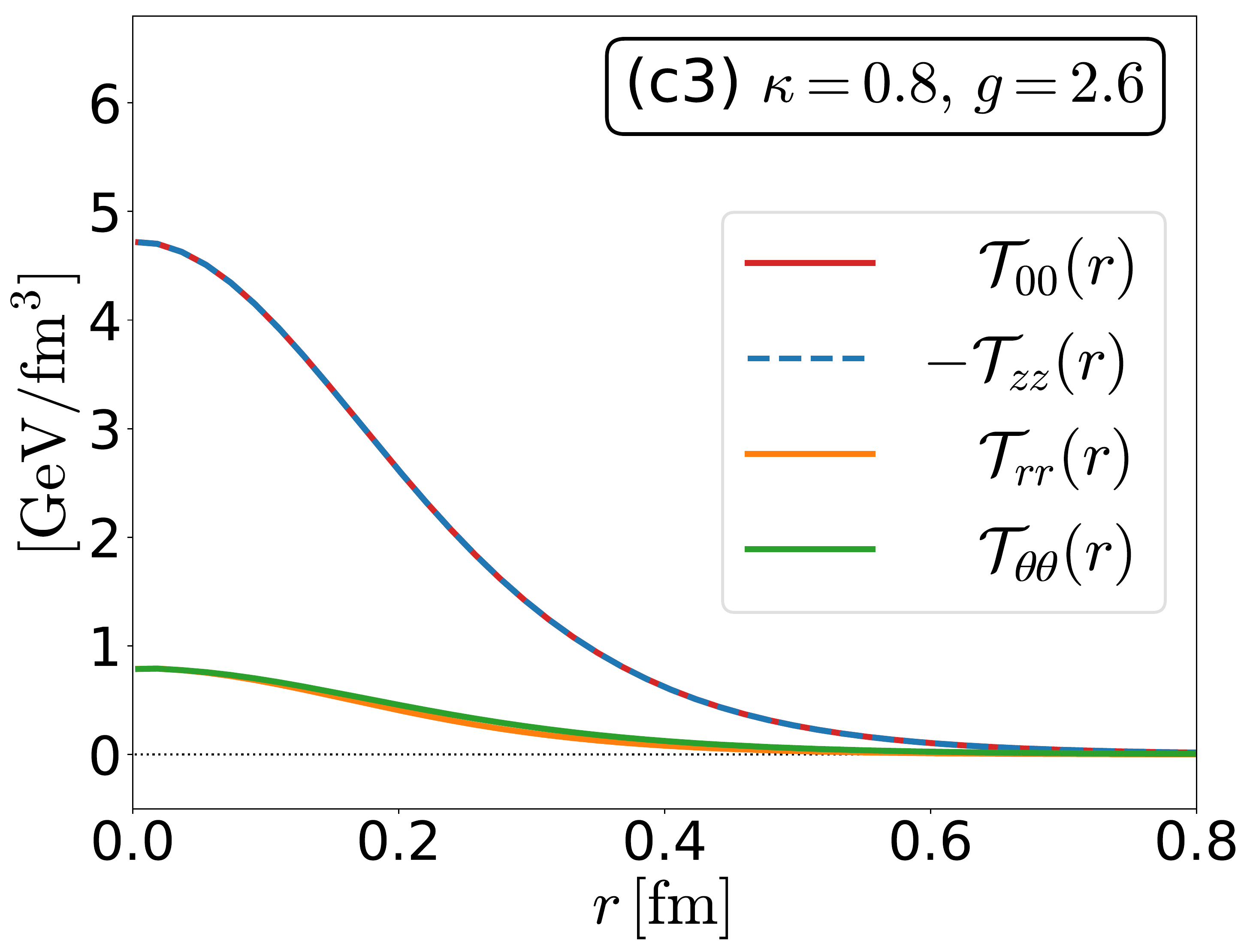}
 \\
 \includegraphics[width=0.325\textwidth]{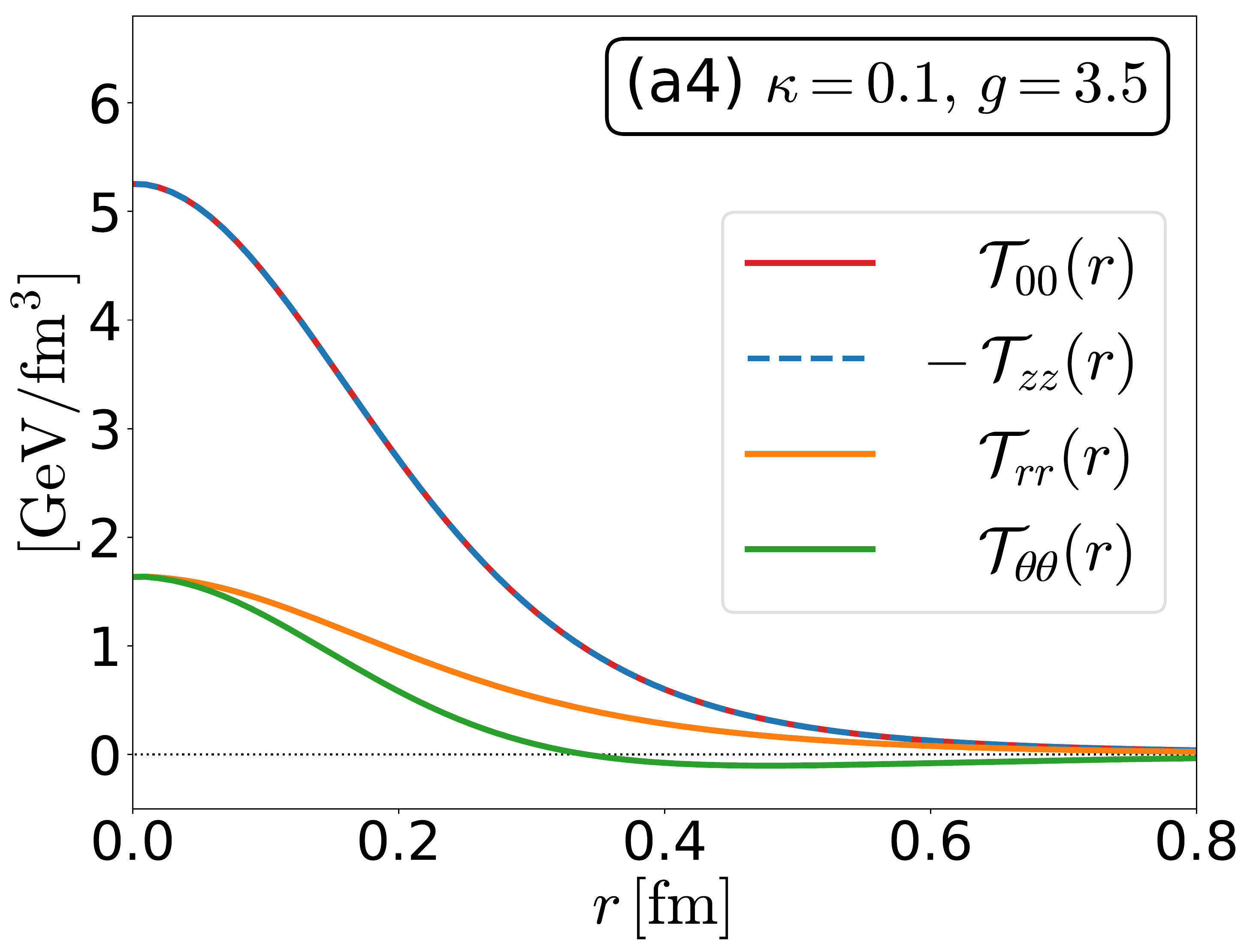}
 \includegraphics[width=0.325\textwidth]{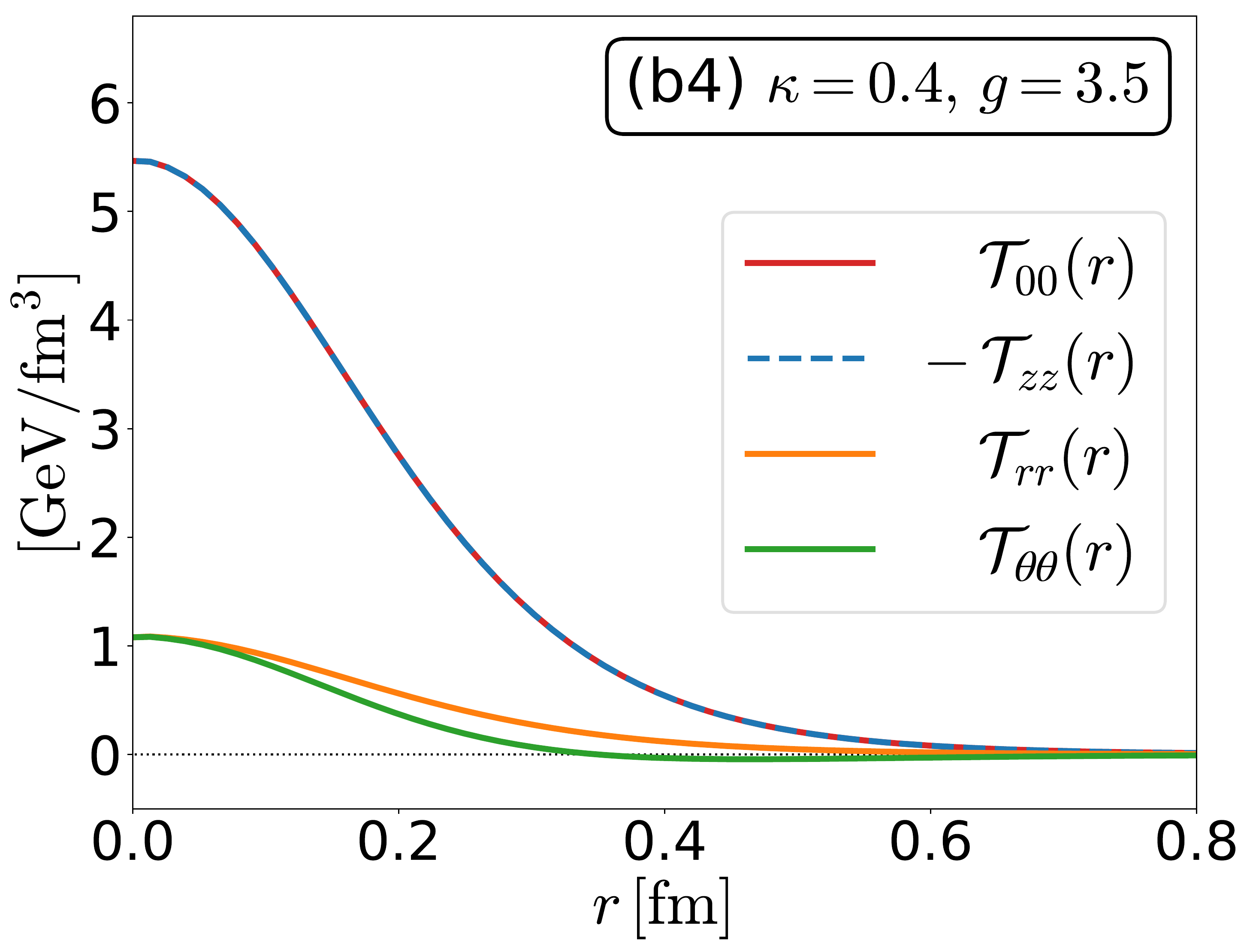}
 \includegraphics[width=0.325\textwidth]{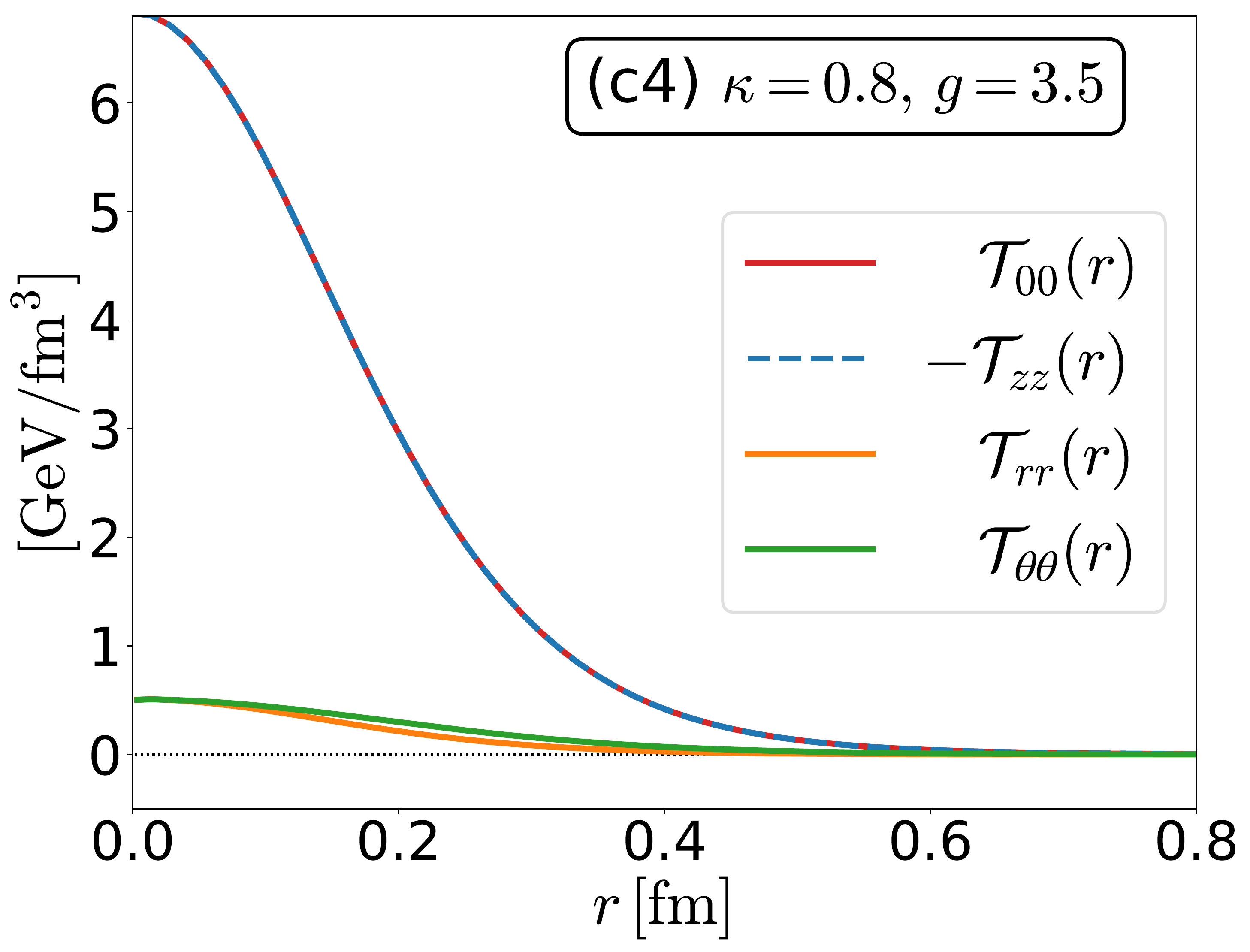}
 \caption{
 EMT distribution on the mid-plane of the magnetic vortex with $R=0.92$~fm
 for various combinations of $\kappa$ and $g$.
 The left, middle, and right panels show the results for
 $\kappa=0.1$, $0.4$ and $0.8$, respectively, 
 while the value of $g$ is $g=0.8,\,1,7,\,2.6$, and $3.5$
 from the top to the bottom.
 }
 \label{fig:result_fin_092}
\end{figure}

\begin{table}
 \centering
 
 \begin{tabular}{|c|c|c|c|c|} \hline
  \multicolumn{2}{|c|}{\backslashbox{$g$}{$\kappa$}} & 0.1 & 0.4 & 0.8\\ \hline
  0.8 & $\xi_\chi\,[\mathrm{fm}]$ & 4.57 & 1.45 & 0.84 \\ \cline{2-5}
  & $\xi_A\,[\mathrm{fm}]$ & 0.65 & 0.82 & 0.94 \\ \hline
  1.7 & $\xi_\chi\,[\mathrm{fm}]$ & 2.15 & 0.68 & 0.39 \\ \cline{2-5}
  & $\xi_A\,[\mathrm{fm}]$ & 0.30 & 0.39 & 0.45 \\ \hline
  2.6 & $\xi_\chi\,[\mathrm{fm}]$ & 1.41 & 0.45 & 0.26 \\ \cline{2-5}
  & $\xi_A\,[\mathrm{fm}]$ & 0.20 & 0.25 & 0.29 \\ \hline
  3.5 & $\xi_\chi\,[\mathrm{fm}]$ & 1.04 & 0.33 & 0.19 \\ \cline{2-5}
  & $\xi_A\,[\mathrm{fm}]$ & 0.15 & 0.19 & 0.22 \\ \hline
 \end{tabular}
 
 \caption{
 Parameters $\kappa$ and $g$ of the numerical analysis in
 Fig.~\ref{fig:result_fin_092}, and the corresponding values of 
 the correlation lengths $\xi_\chi$ and $\xi_A$.
 }
 \label{table:fin_c}
 
\end{table}

We first fix the length of the vortex to be $R=0.92$~fm,
the largest length of the flux tube in Ref.~\cite{Yanagihara:2018qqg},
and study the $\kappa$ and $g$ dependence of the EMT distribution
on the mid-plane.
Shown in Fig.~\ref{fig:result_fin_092} are
the EMT distribution on the mid-plane
for various combinations of $\kappa$ and $g$.
The values of $\kappa$ and $g$ 
increase along right and lower directions, respectively.
The correlation lengths, $\xi_\chi$ and $\xi_A$, corresponding to
each panel are shown in Table~\ref{table:fin_c}.
With fixed $\kappa$, $\xi_\chi$ and $\xi_A$ are monotonically
decreasing as $g$ becomes larger as in Eq.~(\ref{eq:kappa}).
The effect of boundaries thus becomes smaller as $g$ becomes larger.
By increasing $\kappa$ with fixed $g$, on the other hand,
$\xi_\chi$ increases but $\xi_A$ is decreases.
This behavior comes from the $\kappa$ dependence of $v$
in Eq.~(\ref{eq:v}).

From Fig.~\ref{fig:result_fin_092},
one finds that the difference between 
$\mathcal{T}_{rr}(r)$ and $\mathcal{T}_{\theta\theta}(r)$ tends to decrease
as $g$ becomes smaller.
In particular, one sees that these channels are almost degenerated
in the upper two rows, while these channels have a clear separation
from $\mathcal{T}_{00}(r)$ and $\mathcal{T}_{zz}(r)$.
This result suggests that the degeneracy between 
$\mathcal{T}_{rr}(r)$ and $\mathcal{T}_{\theta\theta}(r)$
and their separation from $\mathcal{T}_{00}(r)$ and $\mathcal{T}_{zz}(r)$
observed in Ref.~\cite{Yanagihara:2018qqg} can be described by
the effects of the boundaries.

\begin{figure}
 \centering
 \includegraphics[width=0.48\textwidth]{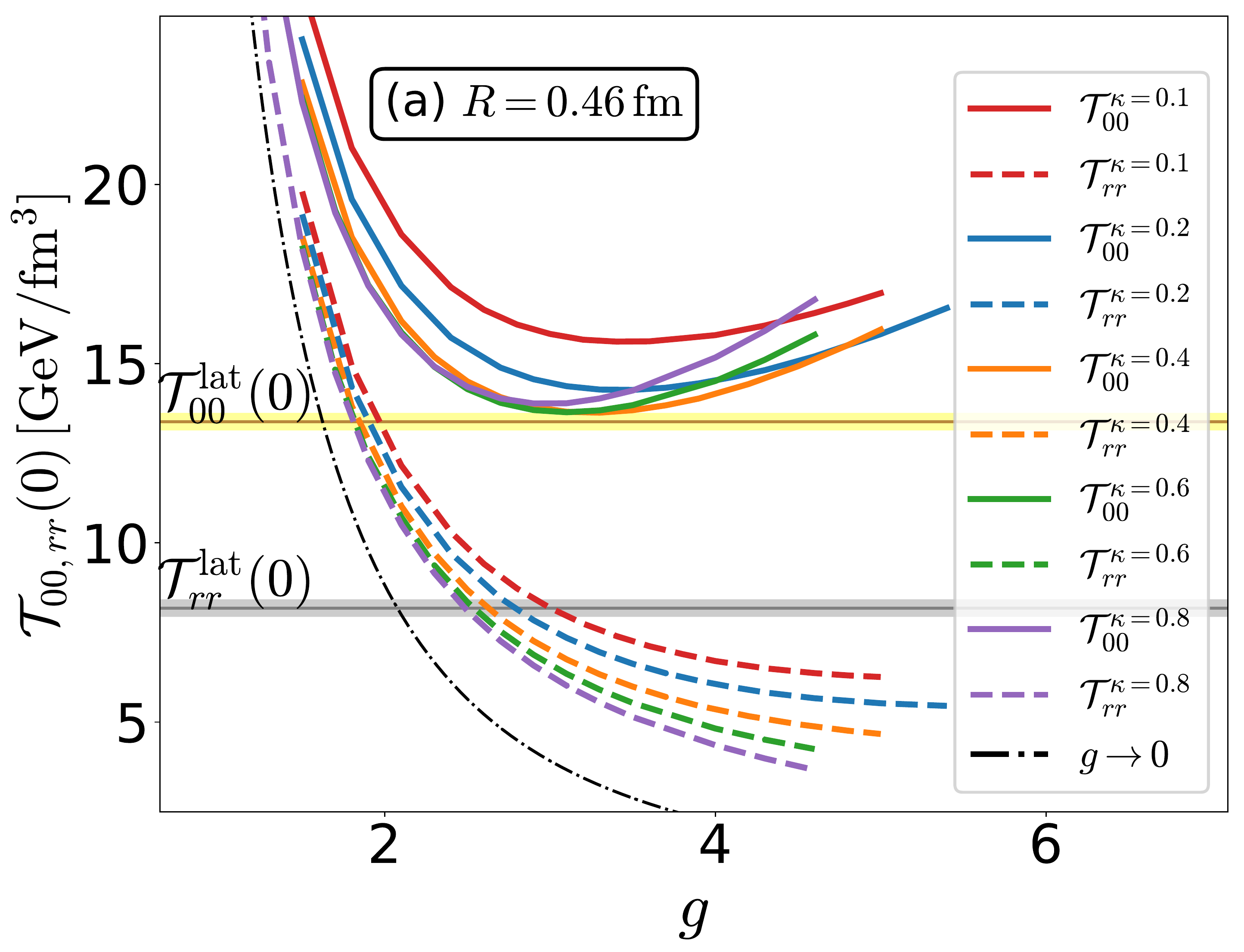}
 \includegraphics[width=0.48\textwidth]{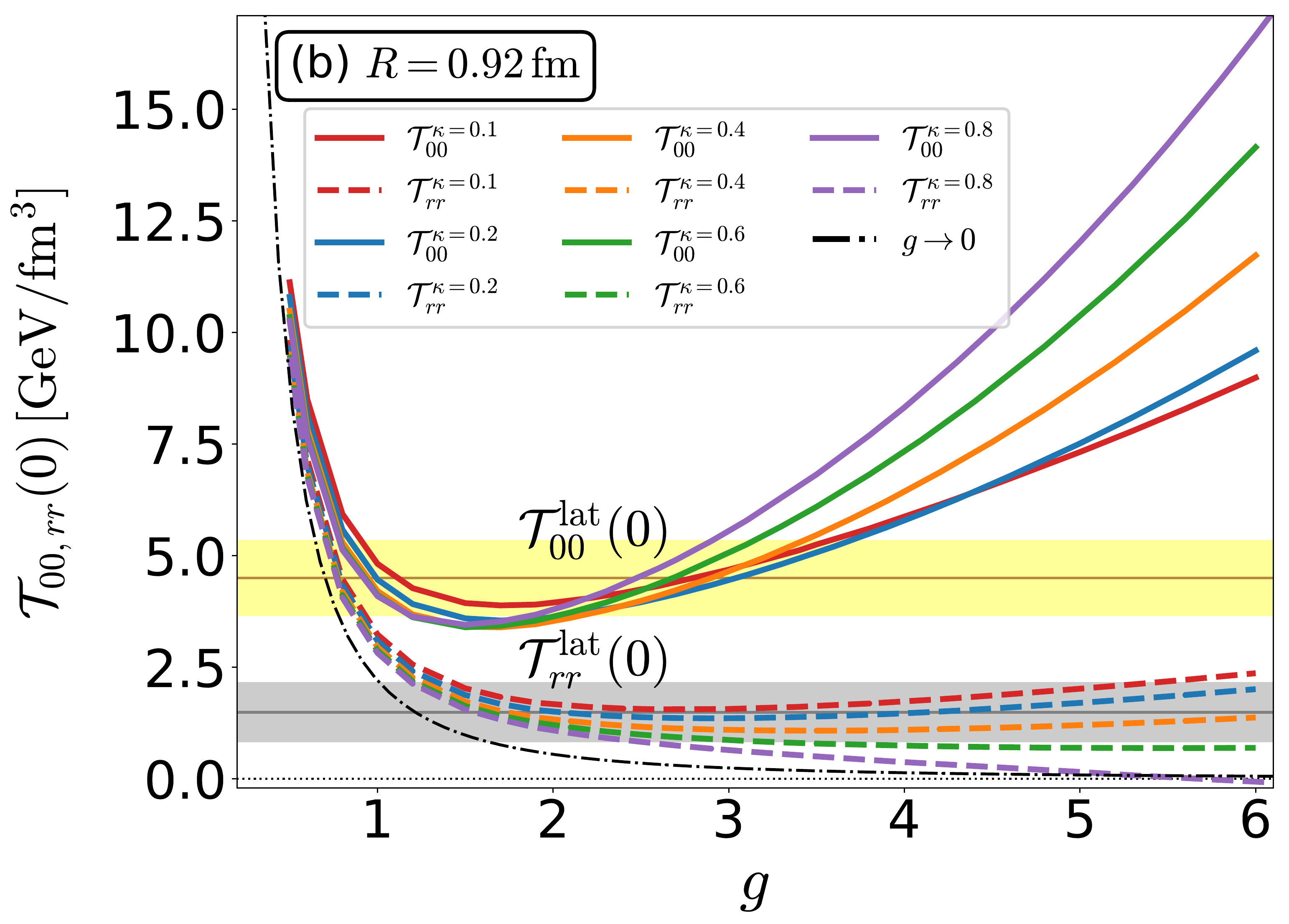}
 \caption{
 Dependences of $\mathcal{T}_{00}(0)$ and $\mathcal{T}_{rr}(0)$
 on $g$ for several values of $\kappa$ at $R=0.46$~fm (left)
 and $R=0.92$~fm (right).
 The dash-dotted line shows the contribution from the gauge field
 $\vec{A}^D(r,z)$,
 Eq.~(\ref{eq:coul_behave}).
 }
 \label{fig:int}
\end{figure}
  
\begin{figure}
 \centering
 \includegraphics[width=0.48\textwidth]{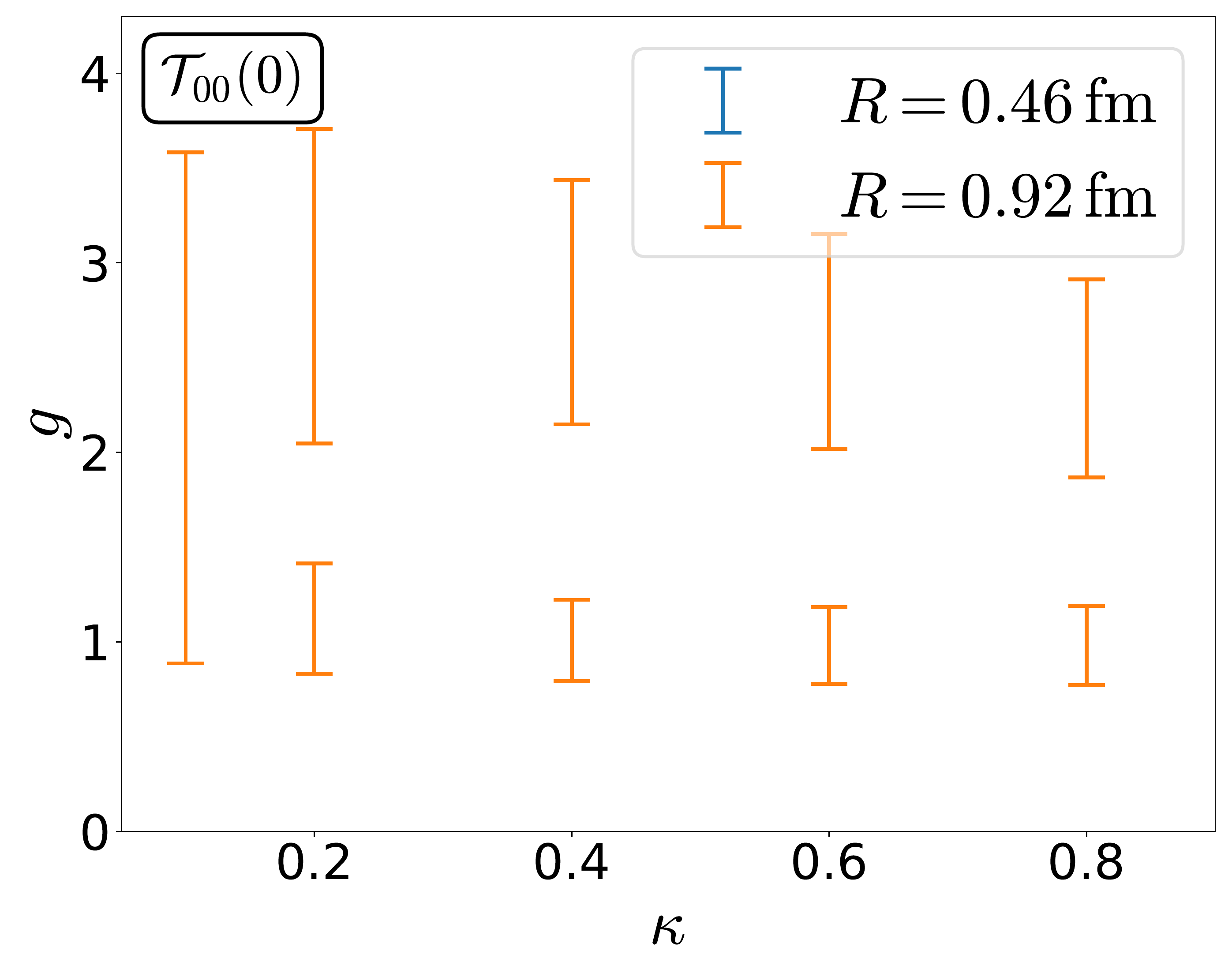}
 \includegraphics[width=0.48\textwidth]{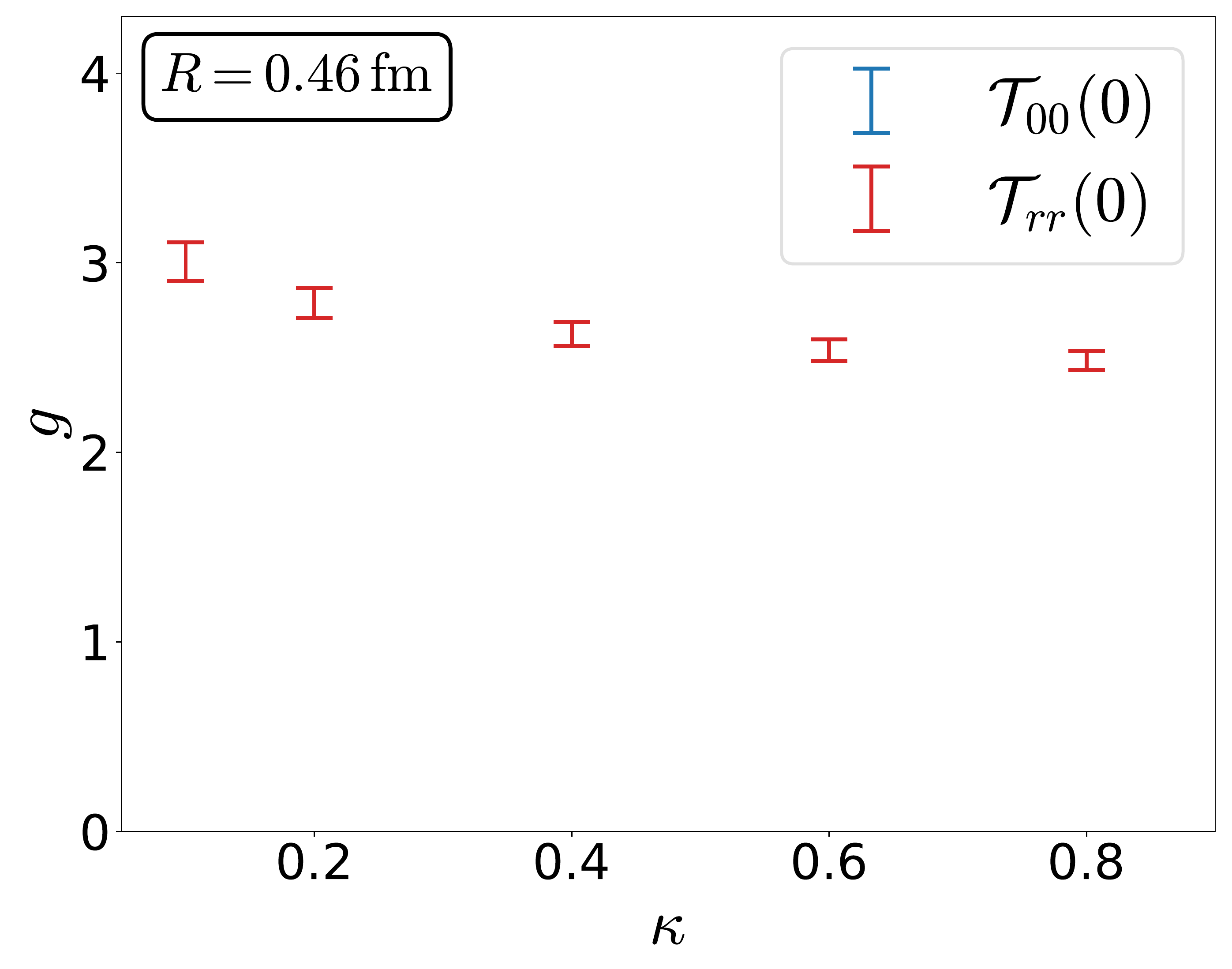}
 \caption{
 (Left) Range of the parameter $g$ which gives the value of 
 $\mathcal{T}_{00}(0)$ consistent with the lattice result
 in Ref.~\cite{Yanagihara:2018qqg}.
 (Right) Range of the parameter $g$ which gives the values of 
 $\mathcal{T}_{00}(0)$ and $\mathcal{T}_{rr}(0)$
 consistent with the lattice result
 in Ref.~\cite{Yanagihara:2018qqg} at $R=0.46$ fm.
 }
 \label{fig:gk}
\end{figure}

\begin{table}
 \centering
 \begin{tabular}{c|cc}
  \hline
  $R$~[fm] & $\mathcal{T}_{00}(0)~{\rm [GeV/fm^3]}$ & $\mathcal{T}_{rr}(0)~{\rm [GeV/fm^3]}$ \\
  \hline
  0.46  &  13.4 (2)  & 8.2 (2) \\
  0.92  &  \phantom{1}4.5 (9) & 1.5 (7) \\
  \hline
 \end{tabular}
 
 \caption{
 The values of $\mathcal{T}_{00}(0)$ and $\mathcal{T}_{rr}(0)$ 
 in Ref.~\cite{Yanagihara:2018qqg} for flux tubes with the length
 $R=0.46$ and $0.92$~fm.
 }
 \label{table:r=0}
\end{table}

Next, to make the comparison between the vortex in the AH model and
lattice results
more quantitatively we focus on the absolute values of
$\mathcal{T}_{00}(r)$ and $\mathcal{T}_{rr}(r)$ at $r=0$.
In Fig.~\ref{fig:int} we show the values of $\mathcal{T}_{00}(0)$
and $\mathcal{T}_{rr}(0)$ as functions of
$g$ for several values of $\kappa$ at $R=0.46$ and $0.92$~fm.
In Fig.~\ref{fig:int}, we also plot Eq.~(\ref{eq:coul_behave})
with $n=1$ by the dash-dotted lines.
When the condition $R\ll\xi_A,\xi_\chi$ is satisfied,
EMT is expected to be dominated by the contribution from
the gauge field $\vec{A}^D(r,z)$ as discussed in Sec.~\ref{sec:mag_vor}.
In this case, which is realized in the small $g$ limit,
$\mathcal{T}_{00}(0)$ and $\mathcal{T}_{rr}(0)$ should 
approach as Eq.~(\ref{eq:coul_behave}).
The figure shows that $\mathcal{T}_{00}(0)$ and $\mathcal{T}_{rr}(0)$
have steep rises corresponding to Eq.~(\ref{eq:coul_behave}) at small $g$.
From the figure one also finds that 
$\mathcal{T}_{00}(0)$ and $\mathcal{T}_{rr}(0)$
go toward degeneracy in this $g$ range.

The horizontal lines in Fig.~\ref{fig:int} show 
the values of $\mathcal{T}_{00}(0)$
and $\mathcal{T}_{rr}(0)$ in Ref.~\cite{Yanagihara:2018qqg}
with the errorbars indicated by the shaded region;
the numerical values of $\mathcal{T}_{00}(0)$ and $\mathcal{T}_{rr}(0)$
are given in Table~\ref{table:r=0}.
We note that the value of $\mathcal{T}_{00}(0)$ used here is obtained
from the average $(\mathcal{T}_{00}(0)+\mathcal{T}_{zz}(0))/2$
in Ref.~\cite{Yanagihara:2018qqg}.
The value of $g$ with fixed $\kappa$ can be constrained
by requiring that $\mathcal{T}_{00}(0)$ of the magnetic vortex
reproduces these lattice result.
In the left panel of Fig.~\ref{fig:gk}, we show the range of
$g$ determined in this way for $R=0.46$ and $0.92$~fm by the bands.
The upper and lower bounds of the bands in the panel are
the values of $g$ at which $\mathcal{T}_{00}(0)$ of the vortex is 
the upper and lower bounds of the errorbar of the lattice result.
We note that there are two ranges of $g$ for 
$R=0.92$~fm and $\kappa\ge0.2$ because of
the non-monotonic behavior of $\mathcal{T}_{00}(0)$ as a function of $g$
as shown in Fig.~\ref{fig:int}.
In the result with $R=0.46$~fm, the value of 
$\mathcal{T}_{00}(0)$ never agrees with the lattice result
in Ref.~\cite{Yanagihara:2018qqg} within the statistical error.
In the left panel of Fig.~\ref{fig:gk}, therefore, 
the data corresponding to
$\mathcal{T}_{00}(0)$ at $R=0.46$~fm are not shown.
This result shows that the vortex solution in the AH model
does not have a parameter set which reproduces the EMT distribution of
the flux tube in SU(3) YM theory for $0.1<\kappa<0.8$.
The same conclusion is also obtained from the 
right panel of Fig.~\ref{fig:gk} that shows the range of $g$
constrained by requiring 
that $\mathcal{T}_{00}(0)$ and $\mathcal{T}_{rr}(0)$
are consistent with the lattice results at $R=0.46$~fm.

\section{Summary}
\label{sec:summary}

In the present study,
motivated by the numerical analysis of the $Q\bar{Q}$ system
in SU(3) YM theory in Ref.~\cite{Yanagihara:2018qqg},
we investigated the EMT distribution around the flux tube and
magnetic vortex.
In Sec.~\ref{sec:stress}, using the momentum conservation 
we have shown that the lattice result on the mid-plane
in Ref.~\cite{Yanagihara:2018qqg}
is qualitatively inconsistent with an assumption of the translational
invariance even at the $Q\bar{Q}$ distance $R=0.92$~fm.
We then employed the AH model in Sec.~\ref{sec:ah} and calculated
the EMT distribution on the mid-plane of the magnetic vortex.
These results are compared with the lattice result on the basis
of the dual superconductor picture~\cite{Suzuki:1988yq,Ball:1987cf,Maedan:1989ju,Suganuma:1993ps,Sasaki:1994sa,Koma:1999sm,Koma:2003gq}.
The results obtained with 
the vortex with finite length suggest
that the degeneracy between $\mathcal{T}_{rr}(r)$ and
$\mathcal{T}_{\theta\theta}(r)$,
and their separation from $\mathcal{T}_{00}(r)$ and $\mathcal{T}_{zz}(r)$,
observed in Ref.~\cite{Yanagihara:2018qqg} 
can be explained qualitatively by the effect of boundaries.
However, from the comparison of the absolute values of 
$\mathcal{T}_{00}(0)$ and $\mathcal{T}_{rr}(0)$,
we have shown that the wide range of the parameters in 
the AH model Eq.~(\ref{eq:L}) cannot reproduce 
the EMT distribution obtained in
Ref.~\cite{Yanagihara:2018qqg} simultaneously,
although a possibility of the existence of a parameter set
in the range $\kappa<0.1$ and $\kappa>0.8$
is not excluded in the present study.

\section*{Acknowledgment}
The authors thank T.~Iritani, M.~Asakawa, and T.~Hatsuda 
for discussions in the early stage of this study.
They also thank N.~Ishii, M.~Koma, K.~Kondo, and A.~Shibata,
H.~Suganuma for discussion.
The authors are also grateful to 
L.~Oxman and M.~Sim\~oes for pointing out 
the errors of the previous version and for making a detailed crosscheck
of the numerical analyses.
MK was supported by JSPS Grant-in-Aid for Scientific Researches
17K05442.

\appendix

\section{Analytic properties}
\label{sec:analytic}

In this appendix, 
we summarize analytic properties of the 
magnetic vortex with an infinite length in the AH model
at the Bogomol'nyi bound 
discussed in Refs.~\cite{Bogomolnyi:1976}.

Throughout this appendix, we use the dimensionless variables $\rho$,
$P(\rho)$, and $Q(\rho)$ defined in Eq.~(\ref{eq:dimless_var}).

The classical vortex solution corresponds to the minimum of
Eq.~(\ref{eq:hatsigma}) 
\begin{align}
 \hat\Sigma[P,Q]
 &=
 \int_0^\infty d\rho \rho
 \Big[ \frac{1}{2\rho^2}
 (\partial_\rho Q)^2 + (\partial_\rho P)^2
 +\frac{P^2Q^2}{\rho^2}
 + \kappa^2 (P^2-1)^2 \Big].
 \label{eq:sigma1}
\end{align}
with the boundary conditions
\begin{align}
 &Q(\rho)\to n , \quad P(\rho)\to 0 \qquad\mbox{for }~ \rho\to0,
 \label{eq:boundary1}
 \\
 &Q(\rho)\to 0 , \quad P(\rho)\to 1 \qquad\mbox{for }~ \rho\to\infty,
 \label{eq:boundary2}
\end{align}
where we allow for an arbitrary winding number $n$ in this Appendix.

At $\kappa=1/\sqrt2$, Eq.~(\ref{eq:sigma1}) is rewritten as
\begin{align}
 \hat\Sigma[P,Q]
 &=
 \int_0^\infty d\rho \rho
 \Big[ \frac12 A_\pm(\rho)^2 + B_\pm(\rho)^2 \Big]
 \mp \int_0^\infty d\rho \partial_\rho ( Q (P^2-1 ) ).
 \label{eq:sigma2}
\end{align}
with 
\begin{align}
 A_\pm(\rho) = \frac{\partial_\rho Q}{\rho} \pm (P^2-1) ,\quad
 B_\pm(\rho) = \partial_\rho P \pm \frac{PQ}{\rho}.
 \label{eq:AB}
\end{align}
Using these variables, the dimensionless EMT
$\hat{ \mathcal T}_{rr}(\rho)$ and $\hat{\mathcal T}_{\theta\theta}(\rho)$
in Eq.~(\ref{eq:dimlessEMT}) are given by 
\begin{align}
 &\hat{\mathcal T}_{rr}(\rho)
 = \frac14 A_+(\rho) A_-(\rho) + \frac12 B_+(\rho) B_-(\rho) ,
 \label{eq:rr}
 \\
 &\hat{\mathcal T}_{\theta\theta}(\rho)
 = -\frac14 A_+(\rho) A_-(\rho) + \frac12 B_+(\rho) B_-(\rho) .
 \label{eq:tt}
\end{align}

The last term in Eq.~(\ref{eq:sigma2}) given by the total derivative
is calculated to be
\begin{align}
 \mp \int_0^\infty d\rho \partial_\rho ( Q (P^2-1 ) ) = \mp n 
\end{align}
with the boundary conditions Eqs.~(\ref{eq:boundary1})
and (\ref{eq:boundary2}) and one obtains 
\begin{align}
 \hat\Sigma[P,Q]
 =
 \int_0^\infty d\rho \rho
 \Big[ \frac12 A_\pm(\rho)^2 + B_\pm(\rho)^2 \Big]
 \mp n.
 \label{eq:sigma3}
\end{align}
Then, the minimum of Eq.~(\ref{eq:sigma3}) is obtained
when $A_\pm(\rho)=0$ and $B_\pm(\rho)=0$ are satisfied for one of the
signs of subscript~\cite{Bogomolnyi:1976}.
Assuming that $P(\rho)$ and $Q(\rho)$ are
monotonic functions of $\rho$ in the vortex solution,
from the boundary conditions one finds that the conditions
$A_+(\rho)=0$ and $B_+(\rho)=0$ are excluded for positive $n$,
and hence the vortex solution must satisfy 
\begin{align}
 A_-(\rho)=B_-(\rho)=0.
 \label{eq:A=B=0}
\end{align}
Substituting Eq.~(\ref{eq:A=B=0}) into Eq.~(\ref{eq:sigma3})
one obtains $\hat\sigma_{\rm AH}(1/\sqrt2)=n$~\cite{Bogomolnyi:1976}.
For $n<0$, $A_+(\rho)=0$ and $B_+(\rho)=0$ are satisfied, and
for an arbitrary $n$ one obtains 
\begin{align}
 \hat\sigma_{\rm AH}(\kappa)  = |n|.
\end{align}
at $\kappa=1/\sqrt2$.

Finally, substituting Eq.~(\ref{eq:A=B=0}) into Eqs.~(\ref{eq:rr}) and
(\ref{eq:tt}),
one easily finds 
that $\mathcal{T}_{rr}(r) = \mathcal{T}_{\theta\theta}(r)=0$
at $\kappa=1/\sqrt2$.

\end{document}